\documentclass{article}
\usepackage{amssymb}
\usepackage{amsfonts}
\usepackage{amsmath}

\begin{document}

\title{Bose-Fermi Equivalence in the Half String Theory}
\author{A Abdurrahman\thanks{%
ababdu@shi.edu} and M Gassem\thanks{%
The Division of Mathematics and Science, South Texas College , 3201 W.
Pecan, McAllen, Texas 78501, E-mail: mgassem@southtexascollege.edu } \\
Department of Physics\\
Shippensburg University of Pennsylvania\\
1871 Old Main Drive\\
Shippensburg, PA 17257\\
USA}
\maketitle

\begin{abstract}
The bosonic representation of the half string ghost in the full string basis
is examined in full. The proof that the comma 3- vertex (matter and ghost)
in the bosonic representation satisfy the Ward-like identities is
established thus completing the proof of the Bose Fermi equivalence in the
half string theory.
\end{abstract}

\section{Introduction}

The work of references \cite%
{Sen-Zwiebach,Rastelli-Sen-Zwiebach,Gross-Taylor-I,Gross-Taylor-II} has
generated much interest in the half string formulation of Witten's theory of
interacting open bosonic strings. An important role in the formulation of
the comma theory is played by the $BRST$ charge $Q$ of the first quantized
theory\cite{A-A-B-II,Abdu-BordesN-II}. In general, the $BRST$ invariance of
the first quantized theory becomes a gauge invariance of the second theory 
\cite{TBanks-Mpeskin,E.Witten-CST,WSiegel,Siegl-Zwiebach}. In the
interacting half string theory, the role of the ghost fields is quite
subtle. The rich structure of the ghost sector of the interacting theory
deserves more consideration. It is possible to study the ghosts using either
bosons or fermions. However, to relate the fermionic ghost formulation to
the bosonic ghost formulation, we have to carry a bosonization procedure
explicitly\cite{E.Witten-CST,Siegl-Zwiebach,Hlousek-Jevicki,ABA}. The
bosonic and the fermionic realization of the ghost fields for the half
string theory have been used before\cite{A-A-B-I,A-A-B-II}, to write down
the ghost part of the three-comma vertex and the proof of equivalence was
only addressed partially\cite{ABA}. Although both formulations give a gauge
invariant theory, their equivalence is not at all transparent. It is the
purpose of this paper to complete the proof of equivalence for both
formulations. The key to the proof as we have seen in the first part of the
proof \cite{ABA} lies on the various identities satisfied by the $G$%
-coefficients that define the comma interacting vertex. To complete the
proof of equivalence\cite{ABA}, we have to show that both realizations of
the comma vertex satisfy the same Ward-like identities.

\section{\protect\bigskip Writing the half string three vertex in the full
string basis}

The half string ghost coordinates $\phi _{j}^{L,R}\left( \sigma \right) $
are defined in the usual way\cite{A-A-B-I,ABA} 
\begin{equation}
\phi _{j}^{L}\left( \sigma \right) =\phi _{r}(\sigma )-\phi _{r}(\frac{\pi }{%
2})\text{, \ \ \ }\phi _{r}^{R}\left( \sigma \right) =\phi _{r}(\pi -\sigma
)-\phi _{r}(\frac{\pi }{2})\text{ \ \ \ \ for \ }0\leq \sigma \leq \frac{\pi 
}{2}  \notag
\end{equation}%
where $\phi _{j}(\sigma )$ is the full string bosonic ghost coordinate and $%
r=1,2,..,N$ (number of strings) 
\begin{equation}
\phi _{j}\left( \sigma \right) =\phi _{0}^{j}+\sqrt{2}\sum_{n=1}^{\infty
}\phi _{n}^{j}\cos n\sigma \text{, \ \ \ for \ }0\leq \sigma \leq \frac{\pi 
}{2}
\end{equation}%
The mode expansion of the left and right pieces reads%
\begin{equation}
\phi _{r}^{h}\left( \sigma \right) =\sqrt{2}\sum_{n=1}^{\infty }\phi
_{2n+1}^{h,r}\cos \left( 2n+1\right) \sigma \text{, \ \ \ for \ }0\leq
\sigma \leq \frac{\pi }{2},\text{ \ \ \ }h=1,2
\end{equation}%
where $h=1,2$ refers to the left $\left( L\right) $ and right $\left(
R\right) $ parts of the string, respectively.

\bigskip\ Like in the matter sector for the half string formulation of
Witten's theory for open bosonic strings, in the half-string approach to the
ghost part of string theory, the elements of the theory are defined by $%
\delta -function$ type overlaps%
\begin{equation}
V_{3}^{HS,\phi }=\exp \left( i\sum_{r=1}^{3}Q_{r}^{\phi }\phi \left( \pi
/2\right) \right) V_{3,0}^{HS,\phi }
\end{equation}%
where%
\begin{equation}
V_{3,0}^{HS,\phi }=\prod\limits_{r=1}^{3}\prod\limits_{\sigma =0}^{\pi
/2}\delta \left( \phi _{r}^{L}\left( \sigma \right) -\phi _{r-1}^{R}\left(
\sigma \right) \right)
\end{equation}%
It is to be understood that $r-1=0\equiv 3$. The factor $Q_{r}^{\phi }$ is
the ghost number insertion at the mid-point which is needed for the $BRST$
invariance of the theory\cite{Gross-Jevicki-I,A-A-B-I} and in this case $%
Q_{1}^{\phi }=Q_{2}^{\phi }=Q_{1}^{3}=1/2$. As we have seen before in the
Hilbert space of the theory, the $\delta -functions$ translate into operator
overlap equations which determine the precise form of the vertex. The ghost
part of the half string vertex in the full string basis has the same
structure as the coordinate one apart from the mid-point insertions 
\begin{equation}
|V_{\phi }^{HS}>=e^{\frac{1}{2}i\sum\limits_{r=1}^{3}\phi ^{r}\left( \pi
/2\right) }V_{\phi }^{HS}\left( \alpha ^{\phi ,1\dag },\alpha ^{\phi ,2\dag
},\alpha ^{\phi ,3\dag }\right) \left\vert 0,N_{ghost}=\frac{3}{2}\right)
_{123}^{\phi }  \label{eqnGhost3vertexHS}
\end{equation}%
where the $\alpha ^{\prime }$s are the bosonic oscillators defined by the
expansion of the bosonized ghost $\left( \phi \left( \sigma \right) ,p^{\phi
}\left( \sigma \right) \right) $ fields and $V_{\phi }^{HS}\left( \alpha
^{\phi ,1\dag },\alpha ^{\phi ,2\dag },\alpha ^{\phi ,3\dag }\right) $ is
the exponential of the quadratic form in the ghost creation operators with
the same structure as the coordinate piece of the vertex

\begin{eqnarray}
V_{\phi }^{HS}\left( \alpha ^{\phi ,1\dag },\alpha ^{\phi ,2\dag },\alpha
^{\phi ,3\dag }\right) &=&\exp \left[ \frac{1}{2}\sum_{r,s=1}^{3}%
\sum_{n,m=1}^{\infty }\alpha _{-n}^{\phi ,r}G_{nm}^{rs}\alpha _{-m}^{\phi
,s}\right. +  \notag \\
&&\left. \sum_{r,s=1}^{3}p_{0}^{\phi ,r}G_{0m}^{rs}\alpha _{-m}^{\phi ,s}+%
\frac{1}{2}\sum_{r,s=1}^{3}p^{\phi ,r}G_{00}^{rs}p_{0}^{\phi ,s}\right]
\end{eqnarray}%
where the the matrix elements $G_{nm}^{rs}$ have been constructed in
previous work \cite{Gassem}.

In the full string, the fermionic ghost overlap equations are%
\begin{eqnarray}
c_{r}\left( \sigma \right) &=&c_{r-1}(\pi -\sigma )\text{\ \ \ }\sigma \in %
\left[ 0,\frac{\pi }{2}\right]  \notag \\
c_{r}\left( \sigma \right) &=&-c_{r+1}(\pi -\sigma )\text{\ \ , \ \ \ }%
\sigma \in \left[ \frac{\pi }{2},\pi \right]
\end{eqnarray}%
and%
\begin{eqnarray}
b_{r}\left( \sigma \right) &=&b_{r-1}(\pi -\sigma )\text{\ \ \ }\sigma \in %
\left[ 0,\frac{\pi }{2}\right]  \notag \\
b_{r}\left( \sigma \right) &=&b_{r+1}(\pi -\sigma )\text{\ \ , \ \ \ }\sigma
\in \left[ \frac{\pi }{2},\pi \right]
\end{eqnarray}%
The proof of the bose-fermi equivalence involves two major obstacles. The
first is to show that the bosonized half string ghosts, (\ref%
{eqnGhost3vertexHS}), satisfy the $c-$ and $b-overlap$ equations displayed
above. To carry out the proof, the authors \cite{ABA}, utilized the
bosonization formulas%
\begin{equation}
c_{+}\left( \sigma \right) =:e^{i\phi _{+}\left( \sigma \right) }:\text{, \
\ \ \ }b_{+}\left( \sigma \right) =:e^{-i\phi _{-}\left( \sigma \right) }:
\end{equation}%
where%
\begin{equation}
\phi \left( \sigma \right) =\frac{1}{2}\left( \phi _{+}\left( \sigma \right)
+\phi _{-}\left( \sigma \right) \right)
\end{equation}%
and%
\begin{equation}
\phi _{\pm }\left( \sigma \right) =\phi _{0}\pm \sigma \left( p_{0}^{\phi }+%
\frac{1}{2}\right) +i\sum_{n=1}^{\infty }\frac{1}{n}\left( \alpha _{n}^{\phi
}e^{\mp in\sigma }-\alpha _{-n}^{\phi }e^{\pm in\sigma }\right)
\end{equation}%
The fermionic ghost coordinates of the bosonic string are anticommuting
fields%
\begin{eqnarray}
c_{\pm }\left( \sigma \right) &=&c(\sigma )\pm i\pi _{b}\left( \sigma
\right) \text{\ \ \ }\sigma \in \left[ 0,\pi \right]  \notag \\
b\pm \left( \sigma \right) &=&\pi _{c}\left( \sigma \right) \pm ib(\sigma )%
\text{\ \ , \ \ \ }\sigma \in \left[ 0,\pi \right]
\end{eqnarray}%
The $c_{+}\left( c_{-}\right) $ are the ghosts for reparametrization of $%
z=\tau +i\sigma $ $\left( \overline{z}=\tau -i\sigma \right) $, respectively
and the $b_{\pm }$ are the corresponding anti-ghosts. These obey the
anticommutation relations%
\begin{eqnarray}
\left\{ c_{n},c_{m}\right\} &=&\left\{ b_{n},b_{m}\right\} =0  \notag \\
\left\{ c_{n},b_{m}\right\} &=&\delta _{n+m\text{ }0}
\end{eqnarray}%
The fermionic half string ghosts are also defined in the usual way\cite%
{A-A-B-II} 
\begin{eqnarray}
c_{r}^{L}\left( \sigma \right) &=&c_{r}(\sigma )-c_{r}(\frac{\pi }{2})\text{%
, \ \ \ }\sigma \in \left[ 0,\frac{\pi }{2}\right]  \notag \\
c_{r}^{R}\left( \sigma \right) &=&c_{r}(\pi -\sigma )-c_{r}(\frac{\pi }{2})%
\text{, \ \ \ }\sigma \in \left[ 0,\frac{\pi }{2}\right]
\end{eqnarray}%
and similar expressions for $b^{L}(\sigma )$ and $b^{R}(\sigma )$.

In the bosonization of the fermionic coordinates, using the standard
procedure (see ref. 17 in ref. 16), it is not obvious that all ingredients
of the theory employing the bosonic field $\phi ^{L}\left( \sigma \right) $
and $\phi ^{R}\left( \sigma \right) $ are equivalent to those constructed
using the original Fermi fields appearing in the left hand side of the above
relations. It has been shown\cite{ABA}, that the ghost vertices in the half
string operator formulation obey the same overlap equations as the fermionic
vertices and are identical. Consequently one is free to use either
formulation of the ghost sector of the theory. In fact this statement was
only partially true, the authors\cite{ABA} failed to establish that the half
string ghost (plus matter) vertex in the bosonic realization of the ghosts
satisfy the same Ward-like identities obeyed by the half string ghost (plus
matter) vertex in the fermionic realization of the ghosts. To complete the
equivalence between the two realization of the half string ghost vertex, we
need to establish the Ward-like identities utilizing the bosonic
representation of the half string ghost as well.

\section{\protect\bigskip The proof of the Ward-like identities}

The Ward-like identities for the Witten vertex matter plus ghost\cite%
{Hlousek-Jevicki} in the fermionic representation are given by%
\begin{equation}
W_{m}^{x+c,r}|V_{W}^{x+c}>=0\text{, \ \ }m=1,2,...  \label{eqnWardIdFermi}
\end{equation}%
where $W_{m}^{x+c,r}$ is the Ward-like operator defined by 
\begin{equation}
W_{m}^{x+c,r}=W_{m}^{x,r}+W_{m}^{c,r}=L_{m}^{x+c,r}+\sum_{s=1}^{3}%
\sum_{n=0}^{\infty }m\widetilde{N}_{mn}^{rs}L_{-n}^{x+c,s}
\end{equation}%
and%
\begin{equation}
|V_{W}^{x+c}>=|V_{W}^{x}>|V_{W}^{c}>
\end{equation}%
The Virasoro generators for both matter and ghost coordinates are given by

\begin{eqnarray}
L_{m}^{x,r} &=&\sum_{k=1}^{\infty }\alpha _{-k}^{r}\alpha _{k+m}^{r}+\frac{1%
}{2}\sum_{k=1}^{m-1}\alpha _{m-k}^{r}\alpha _{k}^{r}+p_{0}^{r}\alpha _{m}^{r}
\label{eqnVirosorXm} \\
L_{0}^{x,r} &=&\frac{1}{2}\left( p_{0}^{r}\right) ^{2}+\sum_{k=1}^{\infty
}\alpha _{-k}^{r}\alpha _{k}^{r},  \label{eqnVirosorX0}
\end{eqnarray}%
and%
\begin{eqnarray}
L_{m}^{c,r} &=&\sum_{k=1}^{\infty }\left[ \left( 2m+k\right)
b_{-k}^{r}c_{k+m}^{r}-\left( m-k\right) c_{-k}^{r}b_{k+m}^{r}\right]  \notag
\\
&&+\sum_{k=1}^{m-1}\left( m+k\right)
b_{m-k}^{r}c_{k}^{r}+mb_{m}^{r}c_{0}^{r}-2mb_{0}^{r}c_{m}^{r} \\
L_{0}^{c,r} &=&\sum_{k=1}^{\infty }k\left(
b_{-k}^{r}c_{k}^{r}-c_{-k}^{r}b_{k}^{r}\right) ,
\end{eqnarray}%
respectively. Here $m$ takes integral values greater than $0$. If the half
string bosonic ghost version of the vertex is equivalent to the fermionic
version then it must obey an identity of the form (\ref{eqnWardIdFermi}) and
the anomaly of the ghost part must cancel the anomaly of the coordinate. Our
job is thus to show that the half string full vertex (i.e., matter plus
ghost) in the bosonic representation satisfies the following identity%
\begin{equation}
W_{m}^{x+\phi ,r}|V_{H}^{x+\phi }>=0\text{, \ \ }m=1,2,...  \label{eqnWardID}
\end{equation}%
where the Ward-like operator in this case is expressed in the bosonic
representation of the ghost%
\begin{equation}
W_{m}^{x+\phi ,r}=W_{m}^{x,r}+W_{m}^{\phi ,r}=L_{m}^{x+\phi
,r}+\sum_{s=1}^{3}\sum_{n=0}^{\infty }m\widetilde{N}_{mn}^{rs}L_{-n}^{x+\phi
,s}
\end{equation}%
As before in the above expression $L_{n}^{x+\phi }=L_{n}^{x}+L_{n}^{\phi }$
and $|V_{H}^{x+\phi }>=|V_{H}^{x}>|V_{H}^{\phi }>$. The Virasoro generators
for the ghost in terms of the bosonic operators are given by%
\begin{eqnarray}
L_{m}^{\phi ,r} &=&\sum_{k=1}^{\infty }\alpha _{-k}^{\phi ,r}\alpha
_{k+m}^{\phi ,r}+\frac{1}{2}\sum_{k=1}^{m-1}\alpha _{m-k}^{\phi ,r}\alpha
_{k}^{\phi ,r}+\left( p_{0}^{\phi ,r}-\frac{3}{2}m\right) \alpha _{m}^{\phi
,r}\text{, }  \label{eqnVirosorGm} \\
L_{0}^{\phi ,r} &=&\frac{1}{2}\left( p_{0}^{\phi ,r}\right) ^{2}-\frac{1}{8}%
+\sum_{k=1}^{\infty }\alpha _{-k}^{\phi ,r}\alpha _{k}^{\phi ,r}\text{,}
\label{eqnVirosorG0}
\end{eqnarray}%
where \ $m>0$. The extra term in (\ref{eqnVirosorGm}), that is, the linear
term in $\alpha _{m}^{\phi ,r}$ arises because of the presence of the $R\phi 
$ term in the action \cite{E.Witten-CST} of the bosonized ghosts%
\begin{equation}
I_{\phi }=\frac{1}{2\pi }\int d^{2}\sigma \left( \partial _{\beta }\phi
\partial ^{\beta }\phi -3iR\phi \right)
\end{equation}%
or alternatively because of the extra linear term in the ghost
energy-momentum tensor%
\begin{equation}
T_{\phi }=\frac{1}{2\pi }\left[ \left( \partial _{\pm }\phi \right) ^{2}-%
\frac{3}{2}\partial _{\pm }\phi \right]
\end{equation}%
The extra term is needed\cite{E.Witten-CST,Gross-Jevicki-I,A-A-B-I} and must
have precisely the coefficient given in (\ref{eqnVirosorGm}), so that $\phi $
can cancel the Virasoro anomaly of the $x^{\mu }$ so that the total Fourier
components of the energy momentum%
\begin{equation}
L_{m}^{r}=L_{m}^{x,r}+L_{m}^{\phi ,r}-\frac{9}{8}\delta _{n0}
\end{equation}%
obey the Virasoro algebra%
\begin{equation}
\left[ L_{m}^{r},L_{n}^{s}\right] =\left( m-n\right) L_{m+n}^{r}
\end{equation}%
which is free of central charge.

We will show that the comma vertex $|V_{HS}^{x+\phi
}>=|V_{HS}^{x}>|V_{HS}^{\phi }>$ indeed satisfies the Ward-like identities
stated in equation (\ref{eqnWardID}). It is more convenient to recast the
comma three-point vertex in the full string oscillator basis\cite{Gassem}.

To express the comma vertex in an exponential form in the creation operators
only we need to commute the annihilation part of the ghost mid-point
insertion $\frac{1}{2}i\sum \phi ^{r}\left( \pi /2\right) $ in (\ref%
{eqnGhost3vertexHS}) through the creation part of the vertex. The normal
mode expansion of the ghost field $\phi \left( \sigma \right) $ is at $\tau
=0$%
\begin{equation}
\phi ^{r}\left( \sigma \right) =\phi _{0}^{r}+\sqrt{2}\sum_{n=1}^{\infty
}\phi _{n}^{r}\cos n\sigma
\end{equation}%
The mid-point of the ghost coordinate is obtained by substituting $\sigma
=\pi /2$ in the above expression 
\begin{equation}
\phi ^{r}\left( \frac{\pi }{2}\right) =\phi _{0}^{r}+i\sum_{n=1}^{\infty
}\lambda _{n}\left( \alpha _{n}^{\phi ,r}-\alpha _{-n}^{\phi ,r}\right)
\end{equation}%
where $\lambda _{n}=\left( n\right) ^{-1}\cos \left( n\pi /2\right) $, $%
n=1,2,3,...$. Let us first consider the first factor in (\ref%
{eqnGhost3vertexHS}) 
\begin{eqnarray}
\exp \left( \frac{1}{2}i\sum\limits_{r=1}^{3}\phi ^{r}\left( \pi /2\right)
\right) &=&N_{1}\exp \left( \frac{1}{2}i\sum\limits_{r=1}^{3}\phi
_{0}^{r}\right) \exp \left( \frac{1}{2}\sum\limits_{r=1}^{3}\sum%
\limits_{n=1}^{\infty }\lambda _{n}\alpha _{-n}^{\phi ,r}\right)  \notag \\
&&\times \exp \left( -\frac{1}{2}\sum\limits_{r=1}^{3}\sum\limits_{n=1}^{%
\infty }\lambda _{n}\alpha _{n}^{\phi ,r}\right)  \label{eqnexpomdpoex}
\end{eqnarray}%
where $N_{1}=\exp \left[ -3\cdot 2^{3}\sum_{n=1}^{\infty }n\lambda _{n}%
\right] $. In obtaining this result we made use of the well known identity%
\begin{equation}
e^{\hat{A}_{1}}e^{\hat{A}_{2}}=e^{\frac{1}{2}\left[ \hat{A}_{1},\hat{A}_{2}%
\right] }e^{\hat{A}_{1}+\hat{A}_{2}}  \label{Eqn IDOPOP}
\end{equation}%
which is valid when $\left[ \hat{A}_{1},\hat{A}_{2}\right] $ is a $C$
number. The next step is to bring the annihilation part of the mid-point
insertions to act on the ghost vacuum. So we need to consider the operator
product%
\begin{equation}
\exp \left( -\frac{1}{2}\sum\limits_{r=1}^{3}\sum\limits_{n=0}^{\infty }%
\widetilde{\lambda }_{n}\alpha _{n}^{\phi ,r}\right) V_{\phi }^{HS}\left(
\alpha ^{\phi ,1\dag },\alpha ^{\phi ,2\dag },\alpha ^{\phi ,3\dag }\right)
\label{eqnopproduct1}
\end{equation}%
where $\widetilde{\lambda }_{n}\equiv \lambda _{n}$ for $n>0$ and $%
\widetilde{\lambda }_{0}=0$. To commute the annihilation operators $\alpha
_{n}^{\phi ,r}$ through the creation operators $a_{-n}^{\phi ,s}$, $%
a_{-m}^{\phi ,s}$, we note that the exponential of the quadratic form is the
Gaussian%
\begin{eqnarray}
V_{\phi }^{HS}\left( \alpha ^{\phi ,1\dag },\alpha ^{\phi ,2\dag },\alpha
^{\phi ,3\dag }\right) &=&\exp \left( \frac{1}{2}\sum_{r,s=1}^{3}%
\sum_{n,m=0}^{\infty }a_{-n}^{\phi ,r}G_{nm}^{rs}a_{-m}^{\phi ,s}\right) 
\notag \\
=\lim_{N\rightarrow \infty }\pi ^{-N/2} &&\left[ \det G\right] ^{-1/2}\int
Dx\exp \left( -\frac{1}{2}\overrightarrow{x}^{T}G^{-1}\overrightarrow{x}%
\right)  \notag \\
&&\times \exp \left( \sum_{s=1}^{3}\sum_{m=0}^{\infty }a_{-m}^{\phi
,s}x_{m}^{s}\right)  \label{eqnGaussianalal}
\end{eqnarray}%
where $Dx=\prod_{r=1}^{3}\prod_{n=0}^{N}dx_{n}^{r}$. Thus we first need to
consider%
\begin{equation}
\exp \left( -\frac{1}{2}\sum\limits_{r=1}^{3}\sum\limits_{n=0}^{\infty }%
\widetilde{\lambda }_{n}\alpha _{n}^{\phi ,r}\right) \exp \left(
\sum_{s=1}^{3}\sum_{m=0}^{\infty }a_{-m}^{\phi ,s}x_{m}^{s}\right)
\end{equation}%
Using the well known identity 
\begin{equation}
e^{\hat{A}_{1}}e^{\hat{A}_{2}}=e^{\left[ \hat{A}_{1},\hat{A}_{2}\right] }e^{%
\hat{A}_{2}}e^{\hat{A}_{1}}
\end{equation}%
which is valid when $\left[ \hat{A}_{1},\hat{A}_{2}\right] $ is a $C$
number, the above product becomes%
\begin{eqnarray}
&&\exp \left( -\frac{1}{2}\sum\limits_{r=1}^{3}\sum\limits_{n=0}^{\infty }%
\widetilde{\lambda }_{n}\alpha _{n}^{\phi ,r}\right) \exp \left(
\sum_{s=1}^{3}\sum_{m=0}^{\infty }a_{-m}^{\phi ,s}x_{m}^{s}\right)
\label{eqnidenlplp} \\
&=&\exp \left[ \sum_{s=1}^{3}\sum_{m=0}^{\infty }\left( -\frac{1}{2}m%
\widetilde{\lambda }_{m}+a_{-m}^{\phi ,s}\right) x_{m}^{s}\right] \exp
\left( -\frac{1}{2}\sum\limits_{r=1}^{3}\sum\limits_{n=0}^{\infty }%
\widetilde{\lambda }_{n}\alpha _{n}^{\phi ,r}\right)  \notag
\end{eqnarray}%
Observe that the operator $\overrightarrow{\alpha }$ translates $%
\overrightarrow{\alpha }^{\dag }$ by $-m\widetilde{\lambda }_{m}/2$. With
the help of the identities in (\ref{eqnGaussianalal}) and (\ref{eqnidenlplp}%
), the operator product in (\ref{eqnopproduct1}) becomes%
\begin{eqnarray}
&&\exp \left[ \frac{1}{2}\sum_{r,s=1}^{3}\sum_{n,m=0}^{\infty }\left( -\frac{%
1}{2}n\widetilde{\lambda }_{n}+a_{-n}^{\phi ,r}\right) G_{nm}^{rs}\left( -%
\frac{1}{2}m\widetilde{\lambda }_{m}+a_{-m}^{\phi ,s}\right) \right]  \notag
\\
&&\times \exp \left( -\frac{1}{2}\sum\limits_{r=1}^{3}\sum\limits_{n=1}^{%
\infty }\widetilde{\lambda }_{n}\alpha _{n}^{\phi ,r}\right)
\label{eqnlamdatelonvertexbeV}
\end{eqnarray}%
Since $\alpha _{n}^{\phi ,r}|0,0,0)=0$, the above expression when acting on
the vacuum of the three strings gives%
\begin{eqnarray}
&&\exp \left( -\frac{1}{2}\sum\limits_{r=1}^{3}\sum\limits_{n=0}^{\infty }%
\widetilde{\lambda }_{n}\alpha _{n}^{\phi ,r}\right) V_{\phi }^{HS}\left(
\alpha ^{\phi ,1\dag },\alpha ^{\phi ,2\dag },\alpha ^{\phi ,3\dag }\right)
|0,0,0>_{\phi }=  \notag \\
&&\exp \left[ \frac{1}{2}\sum_{r,s=1}^{3}\sum_{n,m=0}^{\infty }\left( -\frac{%
1}{2}n\widetilde{\lambda }_{n}+a_{-n}^{\phi ,r}\right) G_{nm}^{rs}\left( -%
\frac{1}{2}m\widetilde{\lambda }_{m}+a_{-m}^{\phi ,s}\right) \right]  \notag
\\
|0,0,0 &>&_{\phi }
\end{eqnarray}%
Using the identity%
\begin{equation}
\sum_{r=1}^{3}G_{nm}^{rs}=\frac{\left( -1\right) ^{n+1}}{n}\delta _{nm}
\end{equation}%
and the definition of $\widetilde{\lambda }$, the above expression becomes%
\begin{eqnarray}
&&\exp \left( -\frac{1}{2}\sum\limits_{r=1}^{3}\sum\limits_{n=0}^{\infty }%
\widetilde{\lambda }_{n}\alpha _{n}^{\phi ,r}\right) V_{\phi }^{HS}\left(
\alpha ^{\phi ,1\dag },\alpha ^{\phi ,2\dag },\alpha ^{\phi ,3\dag }\right)
|0,0,0>_{\phi }  \notag \\
&=&N_{2}\exp \left[ \sum_{s=1}^{3}\sum_{m=1}^{\infty }\lambda
_{m}^{s}a_{-m}^{\phi ,s}\right] V_{\phi }^{HS}\left( \alpha ^{\phi ,1\dag
},\alpha ^{\phi ,2\dag },\alpha ^{\phi ,3\dag }\right) |0,0,0>_{\phi }
\label{eqnlipompaconv}
\end{eqnarray}%
where $N_{2}=\exp \left[ 2^{-3}\sum_{r,s=1}^{3}\sum_{n,m=1}^{\infty }\cos
\left( n\pi /2\right) G_{nm}^{rs}\cos \left( m\pi /2\right) \right] $.
Combining equation (\ref{eqnlipompaconv}), (\ref{eqnexpomdpoex}), and (\ref%
{eqnGhost3vertexHS}), and replacing $\lambda _{n}$ by $\left( 1/n\right)
\cos \left( n\pi /2\right) $, we find%
\begin{eqnarray}
|V_{\phi }^{HS} &>&=N\exp \left( \frac{1}{2}i\sum_{r=1}^{3}\phi
_{0}^{r}\right) \exp \left( \sum_{r=1}^{3}\sum_{n=1}^{\infty }\frac{1}{n}%
\cos \left( \frac{n\pi }{2}\right) \alpha _{-n}^{\phi ,r}\right)  \notag \\
&&\times V_{\phi }^{HS}\left( \alpha ^{\phi ,1\dag },\alpha ^{\phi ,2\dag
},\alpha ^{\phi ,3\dag }\right) |0,0,0>_{\phi }
\end{eqnarray}%
where $N=N_{1}N_{2}$ is a constant that can be absorbed in the overall
normalization constant.

Before we could apply the Virasoro to the vertex we need first to commute
the annihilation operators in the Virasoro generators through the ghost
insertions. Thus we need to consider%
\begin{equation}
\alpha _{n}^{\phi ,r}\exp \left( \sum_{r=1}^{3}\sum_{n=1}^{\infty }\frac{1}{n%
}\cos \left( \frac{n\pi }{2}\right) \alpha _{-n}^{\phi ,r}\right) \text{, }%
n=1,2,3,...
\end{equation}%
If we write $\alpha _{n}^{\phi ,r}$ as $\alpha _{n}^{\phi ,r}=\left(
\partial /\partial \rho _{n}^{r}\right) \exp \left(
\sum_{s=1}^{3}\sum_{m=0}^{\infty }\rho _{m}^{s}\alpha _{m}^{s}\right) |_{%
\overrightarrow{\rho }=0}$ and use the operator identity in (\ref{Eqn IDOPOP}%
), then the above expression becomes%
\begin{eqnarray}
\alpha _{n}^{\phi ,r}\exp \left( \sum_{r=1}^{3}\sum_{n=1}^{\infty }\frac{1}{n%
}\cos \left( \frac{n\pi }{2}\right) \alpha _{-n}^{\phi ,r}\right) &=&\exp
\left( \sum_{r=1}^{3}\sum_{n=1}^{\infty }\frac{1}{n}\cos \left( \frac{n\pi }{%
2}\right) \alpha _{-n}^{\phi ,r}\right)  \notag \\
&&\times \left( \alpha _{n}^{\phi r}+\cos \frac{n\pi }{2}\right)
\end{eqnarray}%
where $n=1,2,3,...$. For $p_{0}^{\phi ,r}$, we have 
\begin{equation}
p_{0}^{\phi ,r}\exp \left( \frac{1}{2}i\sum_{s=1}^{3}\phi _{0}^{r}\right)
=\exp \left( \frac{1}{2}i\sum_{s=1}^{3}\phi _{0}^{s}\right) \left(
p_{0}^{\phi ,r}+\frac{1}{2}\right)
\end{equation}%
Thus we see that the effect of commuting the annihilation operators in the
Virasoro generators through the ghost insertions results in a shift of the
annihilation operator in the Virasoro generator by%
\begin{equation}
\alpha _{n}^{\phi ,r}\rightarrow \left( \alpha _{n}^{\phi ,r}+\cos \frac{%
n\pi }{2}\right)  \label{eqnInserID1}
\end{equation}%
for $n=1,2,3,...$, and%
\begin{equation}
p_{0}^{\phi ,r}\rightarrow \left( p_{0}^{\phi ,r}+\frac{1}{2}\right)
\label{eqnInserID2}
\end{equation}%
Notice that this shift is independent of the string index $r$.

To commute the annihilation part of the $\alpha $ operators in the Virasoro
generators $L$, we need to commute $\alpha _{n}^{r}$ through the creation
part of the vertex; that is; we need%
\begin{equation}
\alpha _{k}^{t}V_{\phi }^{HS}\left( \alpha ^{\phi ,1\dag },\alpha ^{\phi
,2\dag },\alpha ^{\phi ,3\dag }\right)
\end{equation}%
The above expression may be written as%
\begin{eqnarray}
&&\alpha _{k}^{t}V_{\phi }^{HS}\left( \alpha ^{\phi ,1\dag },\alpha ^{\phi
,2\dag },\alpha ^{\phi ,3\dag }\right) \\
&=&\alpha _{k}^{t}\exp \left( \frac{1}{2}\sum_{r,s=1}^{3}\sum_{n,m=0}^{%
\infty }a_{-n}^{\phi ,r}G_{nm}^{rs}a_{-m}^{\phi ,s}\right)  \notag \\
&=&\frac{\partial }{\partial \rho _{k}^{t}}\left\{ \exp \left(
\sum\limits_{s=1}^{3}\sum\limits_{m=0}^{\infty }\rho _{m}^{s}\alpha
_{m}^{\phi ,s}\right) \exp \left( \frac{1}{2}\sum_{r,s=1}^{3}\sum_{n,m=0}^{%
\infty }a_{-n}^{\phi ,r}G_{nm}^{rs}a_{-m}^{\phi ,s}\right) \right\} |_{%
\overrightarrow{\rho }=0}  \notag
\end{eqnarray}%
The expression inside the curly brackets is identical to equation (\ref%
{eqnopproduct1}) with $\rho _{m}^{s}$ replacing $-\frac{1}{2}\widetilde{%
\lambda }_{n}$; thus the result can be obtained from (\ref%
{eqnlamdatelonvertexbeV}) with $\rho _{m}^{s}$ replacing $-\frac{1}{2}%
\widetilde{\lambda }_{n}$; hence the above expression becomes 
\begin{eqnarray}
&&\alpha _{k}^{t}V_{\phi }^{HS}\left( \alpha ^{\phi ,1\dag },\alpha ^{\phi
,2\dag },\alpha ^{\phi ,3\dag }\right) =\frac{\partial }{\partial \rho
_{k}^{t}}\exp \left[ \frac{1}{2}\sum_{r,s=1}^{3}\sum_{n,m=0}^{\infty }\left(
\left( n+\delta _{n0}\right) \rho _{n}^{r}+a_{-n}^{\phi ,r}\right) \right. 
\notag \\
&&\left. G_{nm}^{rs}\left( \left( m+\delta _{m0}\right) \rho
_{m}^{s}+a_{-m}^{\phi ,s}\right) \right] \exp \left(
\sum\limits_{s=1}^{3}\sum\limits_{m=0}^{\infty }\rho _{m}^{s}\alpha
_{m}^{\phi ,s}\right) |_{\overrightarrow{\rho }=0}
\label{eqnIMPOIDCALFTbefore}
\end{eqnarray}%
So we have succeeded in commuting the annihilation operator $\alpha
_{n}^{\phi ,r}$ through the creation part of the vertex. Since $\alpha
_{n}^{\phi ,r}$ annihilates the ghost part of the vacuum of the three
strings, then%
\begin{equation}
\exp \left( \sum\limits_{s=1}^{3}\sum\limits_{m=0}^{\infty }\rho
_{m}^{s}\alpha _{m}^{\phi ,s}\right) |_{\overrightarrow{\rho }%
=0}|0>_{123}^{\phi }=|0>_{123}^{\phi }  \notag
\end{equation}%
and equation (\ref{eqnIMPOIDCALFTbefore}) becomes%
\begin{eqnarray}
&&\alpha _{n}^{r}V_{\phi }^{HS}\left( \alpha ^{\phi ,1\dag },\alpha ^{\phi
,2\dag },\alpha ^{\phi ,3\dag }\right) |0>_{123}^{\phi }  \notag \\
&=&\left( n+\delta _{n0}\right)
\sum\limits_{s=1}^{3}\sum\limits_{m=0}^{\infty }G_{nm}^{rs}a_{-m}^{\phi
,s}V_{\phi }^{HS}\left( \alpha ^{\phi ,1\dag },\alpha ^{\phi ,2\dag },\alpha
^{\phi ,3\dag }\right) |0>_{123}^{\phi }  \label{eqnIMPOIDCALFTVF}
\end{eqnarray}%
where $n,m=0,1,2,3,...$. In fact this relation is equivalent to the $\phi $
and $p_{\phi }$ overlaps on $V_{\phi }^{HS}\left( \alpha ^{\phi ,1\dag
},\alpha ^{\phi ,2\dag },\alpha ^{\phi ,3\dag }\right) |0>_{123}^{\phi }$.
This identity will be very useful as we shall see shortly in commuting all
the annihilation operators $\alpha ^{\phi \prime }$s in the Virasoro
generators $L_{n}$ through $V_{\phi }^{HS}\left( \alpha ^{\phi ,1\dag
},\alpha ^{\phi ,2\dag },\alpha ^{\phi ,3\dag }\right) $.

Let us first commute $L_{m}^{\phi ,r}$ through the ghost insertions. Using
the identities in (\ref{eqnInserID1}) and (\ref{eqnInserID2}), we find

\begin{eqnarray}
&&L_{m}^{\phi ,r}\exp \left( \frac{1}{2}i\sum_{r=1}^{3}\phi _{0}^{r}\right)
\exp \left( \sum_{r=1}^{3}\sum_{n=1}^{\infty }\frac{1}{n}\cos \left( \frac{%
n\pi }{2}\right) \alpha _{-n}^{\phi ,r}\right)  \notag \\
&=&\exp \left( \frac{1}{2}i\sum_{r=1}^{3}\phi _{0}^{r}\right) \exp \left(
\sum_{r=1}^{3}\sum_{n=1}^{\infty }\frac{1}{n}\cos \left( \frac{n\pi }{2}%
\right) \alpha _{-n}^{\phi ,r}\right) \left\{ \left[ L_{m}^{\phi ,r}+\frac{3%
}{2}m\alpha _{m}^{\phi ,r}\right] \right. +  \notag \\
&&\left[ \sum_{k=1}^{\infty }\alpha _{-k}^{\phi ,r}\cos \frac{\left(
k+m\right) \pi }{2}+\sum_{k=1}^{m-1}\alpha _{k}^{r,\phi }\cos \frac{\left(
m-k\right) \pi }{2}+p_{0}^{\phi ,r}\cos \frac{m\pi }{2}+\right.
\label{eqnHasla1} \\
&&\left. \left( \frac{1-3m}{2}\right) \alpha _{m}^{r,\phi }\right] +\left. %
\left[ \frac{1}{2}\sum_{k=1}^{m-1}\cos \pi \frac{m-k}{2}\cos \frac{k\pi }{2}%
+\left( \frac{1}{2}-\frac{3}{2}m\right) \cos \frac{m\pi }{2}\right] \right\}
\notag
\end{eqnarray}%
where $m=1,2,...$. In obtaining the above result we made use of the fact
that $\sum_{k=1}^{m-1}\alpha _{m-k}^{\phi ,r}\cos \frac{k\pi }{2}=\frac{1}{2}%
\sum_{k=1}^{m-1}\alpha _{k}^{r,\phi }\cos \pi \frac{m-k}{2}$. Observe that
the quadratic part\footnote{%
Note that the expression $L_{-m}^{\phi ,r}+\frac{3}{2}m\alpha _{-m}^{\phi
,r} $ is indeed quadratic in the creation-annihilation operators $\alpha
^{\phi } $and $\alpha ^{\phi \dag }$ since the linear term $\frac{3}{2}%
m\alpha _{-m}^{\phi ,r}$ cancels agains the linear term in $L_{-m}^{\phi ,r}$%
.} (i.e., the expression in the first square bracket $L_{m}^{\phi ,r}+\frac{3%
}{2}m\alpha _{m}^{\phi ,r}$) is identical to Virasoro generator for the
orbital part $L_{m}^{x,r}$. Thus its action on $V_{\phi }^{HS}\left( \alpha
^{\phi ,1\dag },\alpha ^{\phi ,2\dag },\alpha ^{\phi ,3\dag }\right) $ $%
|0>_{123}^{\phi }$ is identical to the action of $L_{m}^{x,r}$ on the
coordinate part of the vertex because $V_{\phi }^{HS}\left( \alpha ^{\phi
,1\dag },\alpha ^{\phi ,2\dag },\alpha ^{\phi ,3\dag }\right) $ and $%
V_{x}^{HS}\left( \alpha ^{1\dag },\alpha ^{2\dag },\alpha ^{3\dag }\right) $
has exactly the same structure. The expression in the second square brackets
is linear in oscillators and the expression in the third square brackets has
no oscillators. We still need to compute the effect of passing the
expression in the second brackets through the exponential of the quadratic
form in the ghost creation operators. But before we do that let us first see
the effect of passing $L_{-m}^{\phi ,r}$ through the mid-point insertions.
Since $L_{-m}^{\phi ,r}\equiv L_{m}^{\phi ,r\dag }$, then taking the adjoint
of (\ref{eqnVirosorGm}), we find 
\begin{eqnarray}
&&L_{-m}^{\phi ,r}\exp \left( \frac{1}{2}i\sum_{r=1}^{3}\phi _{0}^{r}\right)
\exp \left( \sum_{r=1}^{3}\sum_{n=1}^{\infty }\frac{1}{n}\cos \left( \frac{%
n\pi }{2}\right) \alpha _{-n}^{\phi ,r}\right)  \notag \\
&=&\exp \left( \frac{1}{2}i\sum_{r=1}^{3}\phi _{0}^{r}\right) \exp \left(
\sum_{r=1}^{3}\sum_{n=1}^{\infty }\frac{1}{n}\cos \left( \frac{n\pi }{2}%
\right) \alpha _{-n}^{\phi ,r}\right) \left\{ \left[ L_{-m}^{\phi ,r}+\frac{3%
}{2}m\alpha _{-m}^{\phi ,r}\right] \right.  \notag \\
&&\left. +\left[ \sum_{k=1}^{\infty }\alpha _{-k-m}^{\phi ,r}\cos \frac{k\pi 
}{2}+\left( \frac{1}{2}-\frac{3}{2}m\right) \alpha _{-m}^{\phi ,r}\right]
\right\}  \label{eqnHasla2}
\end{eqnarray}%
where $m=1,2,...$. Once more the quadratic\footnote{%
See previous footnote.} inside the first square bracket; that is, $%
L_{-m}^{\phi ,r}+\frac{3}{2}m\alpha _{-m}^{\phi ,r}$ is identical to the
orbital part of the Virasoro generator $L_{-m}^{x,r}$ and so its effect on
the exponential of the quadratic form in the ghost creation operators is the
same as the action of $L_{-m}^{x,r}$ on the orbital part of the vertex. The
effect of the expression in the second square brackets needs to be computed
which we shall do later. For the zero mode of the ghost Virasoro generators,
we find

\begin{eqnarray}
&&L_{0}^{\phi ,r}\exp \left( \frac{1}{2}i\sum_{r=1}^{3}\phi _{0}^{r}\right)
\exp \left( \sum_{r=1}^{3}\sum_{n=1}^{\infty }\frac{1}{n}\cos \left( \frac{%
n\pi }{2}\right) \alpha _{-n}^{\phi ,r}\right)  \notag \\
&=&\exp \left( \frac{1}{2}i\sum_{r=1}^{3}\phi _{0}^{r}\right) \exp \left(
\sum_{r=1}^{3}\sum_{n=1}^{\infty }\frac{1}{n}\cos \left( \frac{n\pi }{2}%
\right) \alpha _{-n}^{\phi ,r}\right) \left\{ \left[ L_{0}^{\phi ,r}+\frac{1%
}{8}\right] \right.  \notag \\
&&+\left. \left[ \frac{1}{2}p_{0}^{\phi ,r}+\sum_{k=1}^{\infty }\cos \frac{%
k\pi }{2}\alpha _{-k}^{\phi ,r}\right] +\left[ \frac{1}{2}\cdot \frac{1}{%
2^{2}}-\frac{1}{8}\right] \right\}  \label{eqnHasla0}
\end{eqnarray}%
Once again the expression inside the first square brackets is identical to
that for the orbital zero mode of the Virasoro generator $L_{0}^{x,r}$;
hence its action is the same as that of $L_{0}^{x,r}$ on the orbital part of
the vertex. The second and the third expressions in the second and the third
square brackets have no similar terms in the orbital part $L_{0}^{x,r}$.
Using equations (\ref{eqnHasla1}), (\ref{eqnHasla2}), and (\ref{eqnHasla0})
to commute the ghost part of the Ward-like operator through the mid-point
insertions, we find%
\begin{eqnarray}
&&W_{m}^{\phi ,r}\exp \left( \frac{1}{2}i\sum_{r=1}^{3}\phi _{0}^{r}\right)
\exp \left( \sum_{r=1}^{3}\sum_{n=1}^{\infty }\frac{1}{n}\cos \left( \frac{%
n\pi }{2}\right) \alpha _{-n}^{\phi ,r}\right)  \notag \\
&=&\exp \left( \frac{1}{2}i\sum_{r=1}^{3}\phi _{0}^{r}\right) \exp \left(
\sum_{r=1}^{3}\sum_{n=1}^{\infty }\frac{1}{n}\cos \left( \frac{n\pi }{2}%
\right) \alpha _{-n}^{\phi ,r}\right)  \notag \\
&&\left[ W_{m}^{\phi ,r}\left( 2\right) +W_{m}^{\phi ,r}\left( 1\right)
+\kappa _{m}^{\phi ,r}\left( 1\right) \right]
\end{eqnarray}%
where%
\begin{eqnarray*}
&&W_{m}^{\phi ,r}\left( 2\right) \equiv \sum_{k=1}^{\infty }\alpha
_{-k}^{\phi ,r}\alpha _{k+m}^{r,\phi }+\frac{1}{2}\sum_{k=1}^{m-1}\alpha
_{m-k}^{\phi ,r}\alpha _{k}^{r,\phi }+p_{0}^{\phi ,r}\alpha _{m}^{r,\phi } \\
&&+\sum_{s=1}^{3}\sum_{n=1}^{\infty }m\widetilde{N}_{mn}^{rs}\left[
\sum_{k=1}^{\infty }\alpha _{-k-n}^{\phi ,s}\alpha _{k}^{s,\phi }+\frac{1}{2}%
\sum_{k=1}^{m-1}\alpha _{-k}^{\phi ,s}\alpha _{-n+k}^{\phi ,s}+p_{0}^{\phi
,s}\alpha _{-n}^{\phi ,s}\right] \\
&&+\sum_{s=1}^{3}m\widetilde{N}_{m0}^{rs}\left[ \frac{1}{2}\left(
p_{0}^{\phi ,s}\right) ^{2}+\sum_{k=1}^{\infty }\alpha _{-k}^{\phi ,s}\alpha
_{k}^{\phi ,s}\right] \\
&=&\left[ L_{m}^{\phi ,r}+\frac{3}{2}m\alpha _{m}^{\phi ,r}\right]
+\sum_{s=1}^{3}\sum_{n=1}^{\infty }m\widetilde{N}_{mn}^{rs}\left[
L_{-n}^{\phi ,r}+\frac{3}{2}n\alpha _{-n}^{\phi ,r}\right] \\
&&+\sum_{s=1}^{3}m\widetilde{N}_{m0}^{rs}\left[ L_{0}^{\phi ,s}+\frac{1}{8}%
\right]
\end{eqnarray*}

\begin{eqnarray}
&&W_{m}^{\phi ,r}\left( 1\right) \equiv \left[ \sum_{k=1}^{\infty }\alpha
_{-k}^{\phi ,r}\cos \pi \frac{k+m}{2}+\sum_{k=1}^{m-1}\alpha _{k}^{r,\phi
}\cos \pi \frac{m-k}{2}+p_{0}^{\phi ,r}\cos \frac{m\pi }{2}\right.  \notag \\
&&\left. +\left( \frac{1}{2}-\frac{3}{2}m\right) \alpha _{m}^{r,\phi }\right]
+\sum_{s=1}^{3}\sum_{n=1}^{\infty }m\widetilde{N}_{mn}^{rs}\left[
\sum_{k=1}^{\infty }\alpha _{-k-n}^{\phi ,s}\cos \frac{k\pi }{2}\right. 
\notag \\
&&\left. +\left( \frac{1}{2}-\frac{3}{2}n\right) \alpha _{-n}^{\phi ,s}%
\right] +\sum_{s=1}^{3}m\widetilde{N}_{m0}^{rs}\left[ \frac{1}{2}p_{0}^{\phi
,s}+\sum_{k=1}^{\infty }\cos \frac{k\pi }{2}\alpha _{-k}^{\phi ,s}\right] ,
\end{eqnarray}%
and%
\begin{eqnarray}
\kappa _{m}^{\phi ,r}\left( 1\right) &\equiv &\left[ \frac{1}{2}%
\sum_{k=1}^{m-1}\cos \pi \frac{m-k}{2}\cos \frac{k\pi }{2}+\left( \frac{1}{2}%
-\frac{3}{2}m\right) \cos \frac{m\pi }{2}\right]  \notag \\
+\left[ \frac{1}{2}\cdot \frac{1}{2^{2}}-\frac{1}{8}\right] &=&\text{\ }%
\frac{1+\left( -1\right) ^{m}}{2}\frac{5}{2}\frac{m}{2}\left( -1\right)
^{m/2}  \label{eqnk(1)anomaly}
\end{eqnarray}%
It is important to notice that $W_{m}^{\phi ,r}\left( 2\right) $ has
precisely the same structure as the orbital part $W_{m}^{x,r}$ of the
Ward-like operator. Thus its action on $V_{\phi }^{HS}\left( \alpha ^{\phi
,1\dag },\alpha ^{\phi ,2\dag },\alpha ^{\phi ,3\dag }\right) $ $%
|0,0,0)_{\phi }$ is identical to the action of the orbital part of the
Ward-like operator on the orbital part of the vertex. The action of $%
W_{m}^{x,r}$ on the orbital part of the vertex may be computed with the help
of the results obtained in \cite{A-A-B-I}. Thus, the anomaly of the
quadratic part $W_{m}^{\phi ,r}\left( 2\right) $ is%
\begin{equation}
\kappa _{m}^{\phi ,r}\left( 2\right) =-\frac{1}{2}\sum_{k=1}^{m-1}k\left(
m-k\right) G_{m-k\text{ }k}^{rr}
\end{equation}%
We have seen in appendix B, that $G_{nm}^{rr}$ vanish for all $n+m=odd$
(this is a consequence of the cyclic property of the $G$ coefficients);
hence the above expression reduces to%
\begin{equation}
\kappa _{m}^{\phi ,r}\left( 2\right) =\frac{1+\left( -1\right) ^{m}}{2}\left[
-\frac{1}{2}\sum_{k=1}^{m-1}k\left( m-k\right) G_{m-k\text{ }k}^{rr}\right]
\label{eqnK _m(2)}
\end{equation}%
The closed form of the finite sum may be obtained by considering the first
few values for $m$. For $m=2$, we have%
\begin{equation}
\kappa _{2}^{\phi ,r}\left( 2\right) =-\frac{1}{2}G_{11}^{rr}
\end{equation}%
From ref. \cite{Gassem}, we have%
\begin{equation}
G_{11}^{rr}=-\frac{a_{1}b_{1}}{3}-\frac{1}{\pi }\sqrt{\frac{1}{3}}\left(
a_{1}\widetilde{E}_{1}^{b}-b_{1}\widetilde{E}_{1}^{a}\right)
\end{equation}%
The explicit values of $\widetilde{E}_{1}^{a}$ and $\widetilde{E}_{1}^{b}$
were considered by many authors\cite%
{Gross-Jevicki-II,lousek-Jevicki,ABA,BAA-N,Gassem}. Thus using these values
and the values of $a_{1}=2/3$ and $b_{1}=4/3$, we find%
\begin{equation}
G_{11}^{rr}=-\frac{5}{3^{3}}  \label{eqn Grr11}
\end{equation}%
and so we find%
\begin{equation}
\kappa _{2}^{\phi ,r}\left( 2\right) =-\frac{1}{2}\left[ -\frac{5}{3^{3}}%
\right] =\frac{1}{2}\left( -\frac{5}{3^{3}}\right) \frac{2}{2}\left(
-1\right) ^{2/2}
\end{equation}%
For $m=4$, equation (\ref{eqnK _m(2)}) gives%
\begin{equation}
\kappa _{4}^{\phi ,r}\left( 2\right) =-\frac{1}{2}\left[ 3G_{3\text{ }%
1}^{rr}+4G_{22}^{rr}+3G_{1\text{ }3}^{rr}\right]  \label{eqnK_m=4}
\end{equation}%
The values of $G$'s are given by\cite%
{Gross-Jevicki-II,lousek-Jevicki,ABA,BAA-N,Gassem}%
\begin{equation}
G_{13}^{rr}=G_{31}^{rr}=\frac{2^{5}}{3^{6}}
\end{equation}%
and%
\begin{equation}
G_{22}^{rr}=\frac{2a_{2}b_{2}}{12}+\frac{1}{\pi }\sqrt{\frac{1}{3}}\left[
a_{2}\widetilde{S}_{2}^{b}-b_{2}\widetilde{S}_{2}^{a}\right]
\label{eqnG22need}
\end{equation}%
where

\begin{equation}
{}\widetilde{S}_{2}^{a}=\left[ \widetilde{S}_{1}^{a}-\frac{2}{9}\sqrt{3}\pi %
\right] \frac{3}{2}a_{2}-\frac{1}{3}\sqrt{3}\pi \left( -a_{1}a_{1}+\frac{1}{2%
}a_{2}\right) -\frac{1}{4}\sqrt{3}\pi \left( \frac{1}{2}a_{1}\text{ }\right) 
\text{\ \ }
\end{equation}

\begin{equation}
{}{}\widetilde{S}_{2}^{b}=\left[ \widetilde{S}_{1}^{b}-\frac{4}{9}\sqrt{3}%
\pi \right] \frac{3}{4}b_{2}-\frac{1}{3}\sqrt{3}\pi \left( -b_{1}b_{1}+\frac{%
1}{2}b_{2}\right) -\frac{3}{8}\sqrt{3}\pi \left( \frac{1}{2}b_{1}\right) 
\text{\ }
\end{equation}%
Using the explicit values of the $a$'s, $b$'s and $\widetilde{S}_{1}^{a}$,
we find\cite{Gross-Jevicki-II,lousek-Jevicki,ABA,BAA-N,Gassem}%
\begin{equation}
{}\widetilde{S}_{2}^{a}=-\frac{1}{36}\sqrt{3}\pi \left( 4\ln 2-4\ln
3+1\right)
\end{equation}%
and%
\begin{equation}
{}{}\widetilde{S}_{2}^{b}=-\frac{1}{36}\sqrt{3}\pi \left( 16\ln 2-16\ln
3+5\right) \text{\ }
\end{equation}%
and so equation (\ref{eqnG22need}) yields%
\begin{equation}
G_{22}^{rr}=\frac{13}{2\cdot 3^{5}}
\end{equation}%
Now substituting the values of $G_{13}^{rr}$, $G_{31}^{rr}$ and $G_{22}^{rr}$
into (\ref{eqnK_m=4}), we find%
\begin{equation}
\kappa _{4}^{\phi ,r}\left( 2\right) =-\frac{1}{2}\left[ \frac{10}{27}\right]
=\frac{1}{2}\left( -\frac{5}{3^{3}}\right) \frac{4}{2}\left( -1\right) ^{4/2}
\end{equation}%
Continuing this way we see that for $m=even$, the finite sum in (\ref{eqnK
_m(2)}) has the value%
\begin{equation}
\sum_{k=1}^{m-1}k\left( m-k\right) G_{m-k\text{ }k}^{rr}=-\left( -\frac{5}{%
3^{3}}\right) \frac{m}{2}\left( -1\right) ^{m/2}\text{ \ \ \ \ \ \ }
\end{equation}%
Thus we obtain%
\begin{equation}
\kappa _{m}^{\phi ,r}\left( 2\right) =\frac{1+\left( -1\right) ^{m}}{2}\left[
\frac{1}{2}\left( -\frac{5}{3^{3}}\right) \frac{m}{2}\left( -1\right) ^{m/2}%
\right] \text{ }  \label{eqnanomalyQuadrKmPhi}
\end{equation}%
The anomaly for the quadratic coordinate piece has been evaluated by many
authors \cite{Gross-Jevicki-II,Hlousek-Jevicki,ABA,BAA-N,Gassem}%
\begin{equation}
\kappa _{m}^{x,r}\left( 2\right) =\frac{1+\left( -1\right) ^{m}}{2}\left[ 
\frac{D}{2}\left( -\frac{5}{3^{3}}\right) \frac{m}{2}\left( -1\right) ^{m/2}%
\text{ }\right]  \label{eqnanomalyQuadrKmX}
\end{equation}%
Combining equations (\ref{eqnk(1)anomaly}), (\ref{eqnanomalyQuadrKmPhi}),
and (\ref{eqnanomalyQuadrKmX}), we see that the total anomaly 
\begin{equation}
\kappa _{m}^{x,r}\left( 2\right) +\kappa _{m}^{\phi ,r}\left( 2\right)
+\kappa _{m}^{\phi ,r}\left( 1\right) =0
\end{equation}%
in the critical dimension $D=26$. This result provides a nontrivial
consistency check on the validity of the half string theory.

Now we proceed to consider the Linear part of the Ward-like operator. Using
the identity in (\ref{eqnIMPOIDCALFTVF}), we can commute the Ward-like
operator $W_{m}^{\phi ,r}\left( 1\right) $ through $V_{\phi }^{HS}\left(
\alpha ^{\phi ,1\dag },\alpha ^{\phi ,2\dag },\alpha ^{\phi ,3\dag }\right)
> $, skipping some rather simple algebra, we find 
\begin{eqnarray*}
&&W_{m}^{\phi ,r}\left( 1\right) V_{\phi }^{HS}\left( \alpha ^{\phi ,1\dag
},\alpha ^{\phi ,2\dag },\alpha ^{\phi ,3\dag }\right) |0>_{123}^{\phi } \\
&=&\left\{ \sum_{k=1}^{\infty }\alpha _{-k}^{\phi ,r}\cos \pi \frac{k+m}{2}%
+\sum_{k=1}^{m-1}\sum_{s=1}^{3}\sum_{q=0}^{\infty }kG_{kq}^{rs}a_{-q}^{\phi
,s}\cos \pi \frac{m-k}{2}\right. \\
&&+p_{0}^{\phi ,r}\cos \frac{m\pi }{2}+\left( \frac{1}{2}-\frac{3}{2}%
m\right) \sum_{s=1}^{3}\sum_{q=0}^{\infty }mG_{mq}^{rs}a_{-q}^{\phi ,s} \\
&&+\sum_{s=1}^{3}\sum_{n=1}^{\infty }m\widetilde{N}_{mn}^{rs}\left[
\sum_{k=1}^{\infty }\alpha _{-k-n}^{\phi ,s}\cos \frac{k\pi }{2}+\left( 
\frac{1}{2}-\frac{3}{2}n\right) \alpha _{-n}^{\phi ,s}\right] + \\
&&\left. \sum_{s=1}^{3}m\widetilde{N}_{m0}^{rs}\left[ \frac{1}{2}p_{0}^{\phi
,s}+\sum_{k=1}^{\infty }\cos \frac{k\pi }{2}\alpha _{-k}^{\phi ,s}\right]
\right\} V_{\phi }^{HS}\left( \alpha ^{\phi ,1\dag },\alpha ^{\phi ,2\dag
},\alpha ^{\phi ,3\dag }\right) |0>_{123}^{\phi }
\end{eqnarray*}%
Making the identification $\alpha _{0}^{\phi ,s}=p_{0}^{\phi ,s}$ and then
using the fact that%
\begin{equation}
\sum_{q=1}^{m-1}qG_{q0}^{rs}\cos \pi \frac{m-q}{2}=\sum_{q=1}^{m-1}\left(
m-q\right) G_{m-q\text{ }0}^{rs}\cos \pi \frac{q}{2}\text{,}
\end{equation}%
the above expression becomes

\begin{eqnarray*}
&&\left\{ \sum_{s=1}^{3}\sum_{k=1}^{\infty }\left( \delta ^{rs}\cos \pi 
\frac{k+m}{2}\right) \alpha _{-k}^{\phi ,s}+\sum_{s=1}^{3}\sum_{k=1}^{\infty
}\left( \sum_{q=1}^{m-1}qG_{qk}^{rs}\cos \pi \frac{m-q}{2}\right)
a_{-k}^{\phi ,s}\right. \\
&&+\sum_{s=1}^{3}\sum_{k=1}^{\infty }\left( \frac{1}{2}-\frac{3}{2}m\right)
mG_{mk}^{rs}a_{-k}^{\phi ,s}+\sum_{s=1}^{3}\sum_{k=1}^{\infty }\left( m%
\widetilde{N}_{m0}^{rs}\cos \frac{k\pi }{2}\right) \alpha _{-k}^{\phi ,s} \\
&&+\sum_{s=1}^{3}\sum_{n=1}^{\infty }m\widetilde{N}_{mn}^{rs}\sum_{k=1}^{%
\infty }\alpha _{-k-n}^{\phi ,s}\cos \frac{k\pi }{2}+\sum_{s=1}^{3}%
\sum_{k=1}^{\infty }m\widetilde{N}_{mk}^{rs}\left( \frac{1}{2}-\frac{3}{2}%
k\right) \alpha _{-k}^{\phi ,s} \\
&&+\sum_{s=1}^{3}\left[ \delta ^{rs}\cos \frac{m\pi }{2}+\frac{1}{2}m%
\widetilde{N}_{m0}^{rs}+\left( \frac{1}{2}-\frac{3}{2}m\right)
mG_{m0}^{rs}\right. \\
&&\left. +\sum_{q=1}^{m-1}\left( m-q\right) G_{m-q\text{ }0}^{rs}\cos \pi 
\frac{q}{2}\right] \left. p_{0}^{\phi ,s}\right\} V_{\phi }^{HS}\left(
\alpha ^{\phi ,1\dag },\alpha ^{\phi ,2\dag },\alpha ^{\phi ,3\dag }\right)
\left\vert 0\right\rangle _{123}^{\phi }
\end{eqnarray*}%
If we now let $k+n\longrightarrow q$ in the double sum $\sum_{n=1}^{\infty
}\sum_{k=1}^{\infty }\left( ....\right) $ , so that%
\begin{eqnarray*}
&&\sum_{n=1}^{\infty }m\widetilde{N}_{mn}^{rs}\sum_{k=1}^{\infty }\alpha
_{-k-n}^{\phi ,s}\cos \frac{k\pi }{2} \\
&=&\sum_{n=1}^{\infty }\sum_{k=1+n}^{\infty }m\widetilde{N}_{mn}^{rs}\alpha
_{-k}^{\phi ,s}\cos \pi \frac{k-n}{2}=\sum_{k=1}^{\infty }\sum_{n=1}^{k-1}m%
\widetilde{N}_{mn}^{rs}\alpha _{-k}^{\phi ,s}\cos \pi \frac{k-n}{2}
\end{eqnarray*}%
(to see the last equality you only need to expand both sides and compare
terms), we obtain%
\begin{eqnarray}
&&W_{m}^{\phi ,r}\left( 1\right) V_{\phi }^{HS}\left( \alpha ^{\phi ,1\dag
},\alpha ^{\phi ,2\dag },\alpha ^{\phi ,3\dag }\right) \left\vert
0,0,0\right\rangle _{\phi }  \notag \\
&=&\left[ \sum_{s=1}^{3}\sum_{k=1}^{\infty }\Omega _{mk}^{rs}\alpha
_{-k}^{\phi ,s}+\sum_{s=1}^{3}\Omega _{m0}^{rs}\text{ }p_{0}^{\phi ,s}\right]
\notag \\
&&V_{\phi }^{HS}\left( \alpha ^{\phi ,1\dag },\alpha ^{\phi ,2\dag },\alpha
^{\phi ,3\dag }\right) \left\vert 0,0,0\right\rangle _{\phi }
\label{eqnidentoperward}
\end{eqnarray}%
where for $k,m=1,2,3,...$%
\begin{eqnarray}
\Omega _{mk}^{rs} &\equiv &\delta ^{rs}\cos \frac{\left( k+m\right) \pi }{2}%
+\sum_{n=1}^{m-1}nG_{nk}^{rs}\cos \frac{\left( m-n\right) \pi }{2}+\left( 
\frac{1}{2}-\frac{3}{2}m\right) mG_{mk}^{rs}+  \notag \\
&&m\widetilde{N}_{m0}^{rs}\cos \frac{k\pi }{2}+\sum_{n=1}^{k-1}m\widetilde{N}%
_{mn}^{rs}\cos \frac{\left( k-n\right) \pi }{2}+m\widetilde{N}%
_{mk}^{rs}\left( \frac{1}{2}+\frac{3}{2}k\right)  \label{eqnOmegars;km}
\end{eqnarray}%
and for $k=0,m=1,2,3,...$ 
\begin{eqnarray}
\Omega _{m0}^{rs} &\equiv &\delta ^{rs}\cos \frac{m\pi }{2}+\frac{1}{2}m%
\widetilde{N}_{m0}^{rs}+\left( \frac{1}{2}-\frac{3}{2}m\right) mG_{m0}^{rs} 
\notag \\
&&+\sum_{n=1}^{m-1}\left( m-n\right) G_{m-n\text{ }0}^{rs}\cos \pi \frac{n}{2%
}  \label{eqnOmegak=0N}
\end{eqnarray}%
The identities in (\ref{eqnOmegars;km}) and (\ref{eqnOmegak=0N}) are
identical to those obtained Hlousek and Jevicki\cite{Hlousek-Jevicki} for
the full string. In obtaining equation (\ref{eqnidentoperward}) we have
labeled the dummy index \ $q$ as $n$. Thus to establish that%
\begin{equation}
W_{m}^{\phi ,r}\left( 1\right) V_{\phi }^{HS}\left( \alpha ^{\phi ,1\dag
},\alpha ^{\phi ,2\dag },\alpha ^{\phi ,3\dag }\right) |0>_{123}^{\phi }=0%
\text{,}
\end{equation}
we need to prove that%
\begin{equation}
\sum_{s=1}^{3}\sum_{k=1}^{\infty }\Omega _{mk}^{rs}\alpha _{-k}^{\phi
,s}V_{\phi }^{HS}\left( \alpha ^{\phi ,1\dag },\alpha ^{\phi ,2\dag },\alpha
^{\phi ,3\dag }\right) |0>_{123}^{\phi }=0
\end{equation}%
and%
\begin{equation}
\sum_{s=1}^{3}\Omega _{m0}^{rs}p_{0}^{\phi ,s}V_{\phi }^{HS}\left( \alpha
^{\phi ,1\dag },\alpha ^{\phi ,2\dag },\alpha ^{\phi ,3\dag }\right)
|0>_{123}^{\phi }=0
\end{equation}%
We observe that for $k\neq 0$, the states $\alpha _{-k}^{\phi ,s}V_{\phi
}^{HS}\left( \alpha ^{\phi ,1\dag },\alpha ^{\phi ,2\dag },\alpha ^{\phi
,3\dag }\right) |0>_{123}^{\phi }$ are all linearly independent for $s=1,2,3$
and $k=1,2,...$, thus the only way for this part to vanish is for $\Omega
_{mk}^{rs}$ to be identically zero for all values of $s=1,2,3$ and $k=1,2,..$%
. That is we need to show that%
\begin{equation}
\Omega _{mk}^{rs}=0  \label{eqn GeneralIDforOmegars,mk}
\end{equation}%
is true for all values of $r,s=1,2,3$ and $k,m=1,2,..$. For the second part
we need to prove that%
\begin{equation}
\sum_{s=1}^{3}\Omega _{m0}^{rs}p_{0}^{\phi ,s}=0
\end{equation}%
for $m=1,2,..$. Using the conservation of momentum (or more precisely the
ghost number conservation) to eliminate \footnote{%
If we choose to eliminate $p_{0}^{\phi ,1}$or $p_{0}^{\phi ,2}$ instead of $%
p_{0}^{\phi ,3}$ the conclusions remain the same.} $p_{0}^{\phi ,3}$, we see
at once that to establish the above equation we only need to show that the
following identities%
\begin{eqnarray}
\left( \Omega _{m0}^{r1}-\Omega _{m0}^{r3}\right) &=&0  \label{eqnOmega=0I}
\\
\left( \Omega _{m0}^{r2}-\Omega _{m0}^{r3}\right) &=&0  \label{eqnOmega=0II}
\end{eqnarray}%
are satisfied for $r=1,2,3$, $m=1,2,..$. To prove that $\Omega _{mk}^{rs}$
in equation (\ref{eqnOmegars;km}) vanish for all values of $r,s,m,k$ is
quite cumbersome. Unfortunately, it appears to be no short cuts in proving $%
\Omega _{mk}^{rs}=0$ and so we are forced to prove that $\Omega _{mk}^{rs}=0$
by brute force. The identities in (\ref{eqnOmega=0I}) and (\ref{eqnOmega=0II}%
) may be proven with the help of the properties of the coefficients $%
G_{nm}^{rs}$ and $\widetilde{N}_{nm}^{rs}$\cite%
{Gross-Jevicki-II,lousek-Jevicki,ABA,BAA-N,Gassem}. First let us concentrate
on the identities in (\ref{eqnOmega=0I}) and (\ref{eqnOmega=0II}). For $%
m=odd=2k+1>0$, equation (\ref{eqnOmegak=0N}) reduces to%
\begin{eqnarray}
\Omega _{2k+10}^{rs} &=&\frac{1}{2}\left( 2k+1\right) \widetilde{N}%
_{2k+10}^{rs}+\left( \frac{1}{2}-\frac{3}{2}\left( 2k+1\right) \right)
\left( 2k+1\right) G_{2k+10}^{rs}  \notag \\
&&+\sum_{n=1}^{2k}\left( 2k+1-n\right) G_{2k+1-n\text{ }0}^{rs}\cos \pi 
\frac{n}{2}  \label{eqnOmegak=0/m=odd}
\end{eqnarray}%
For $r=s$, we have $G_{odd\text{ }even}^{rr}=\widetilde{N}_{odd\text{ }%
even}^{rr}=0$, and so the above expression vanish since for $n=even$, $%
G_{2k+1-n\text{ }0}^{rs}=0$ and for $n=odd$, $\cos \pi n/2$ vanish. Thus for 
$r=s$, we have%
\begin{equation*}
\Omega _{2k+10}^{rr}=0
\end{equation*}%
It follows from the above identity and (\ref{eqnOmega=0I}) and (\ref%
{eqnOmega=0II}) that to prove the identities in (\ref{eqnOmega=0I}) and (\ref%
{eqnOmega=0II}) for $m=odd=2k+1>0$, we have to establish that 
\begin{equation}
\Omega _{2k+1,0}^{12}=\Omega _{2k+1,0}^{23}=\Omega _{2k+1,0}^{31}=\Omega
_{2k+1,0}^{13}=\Omega _{2k+1,0}^{32}=\Omega _{2k+1,0}^{21}=0
\label{eqnIdentities Odd0r,s}
\end{equation}%
It is not hard to check that the above identities are satisfied for $%
2k+1=1,2,3,..$, by explicit substitution. The proof for a general value of $%
2k+1$, can be established by mathematical induction. We have proved that in
fact these identities are satisfied for all values $2k+1$; however to
present the proof here for all cases will occupy too much space. To
illustrate the proof we will present here the complete proof for one of
them. Let us consider 
\begin{equation}
\Omega _{2k+1,0}^{13}=0
\end{equation}%
From (\ref{eqnOmegak=0/m=odd}) we have%
\begin{eqnarray}
\Omega _{2k+10}^{13} &=&\frac{1}{2}\left( 2k+1\right) \widetilde{N}%
_{2k+10}^{13}+\left( \frac{1}{2}-\frac{3}{2}\left( 2k+1\right) \right)
\left( 2k+1\right) G_{2k+10}^{13}  \notag \\
&&+\sum_{n=1}^{2k}\left( 2k+1-n\right) G_{2k+1-n\text{ }0}^{13}\cos \pi 
\frac{n}{2}
\end{eqnarray}%
From \cite{Gassem,Gross-Jevicki-II}, we have%
\begin{eqnarray}
G_{2k+10}^{13} &=&-\frac{1}{\sqrt{3}}\left( -\right) ^{k}\frac{a_{2k+1}}{2k+1%
} \\
\widetilde{N}_{2k+10}^{13} &=&-\frac{1}{\sqrt{3}}\left( -\right) ^{k}\frac{%
b_{2k+1}}{2k+1}
\end{eqnarray}%
And so the above equation becomes%
\begin{equation}
\Omega _{2k+10}^{13}=\frac{1}{\sqrt{3}}\left( -\right) ^{k}\left[ -\frac{1}{2%
}b_{2k+1}+\left( 3k+1\right) a_{2k+1}-\sum_{m=1}^{k}a_{2k+1-2m}\right]
\label{Omega (13) 2k+10}
\end{equation}%
To proceed further, we need to eliminate either $b$ or $a$ from the above
expression. This can be established with the help of the mixed recursion
relation%
\begin{equation}
\frac{2}{3}b_{n}=\left( -1\right) ^{n}\left[ \left( n+1\right)
a_{n+1}-2na_{n}+\left( n-1\right) a_{n-1}\right]  \label{eqnMixed Recab}
\end{equation}%
which can be established with the help of contour integration\cite{Gassem}.
Thus setting $n=2k+1$ in the recursion relation and substituting for $%
b_{2k+1}$, the above expression in (\ref{Omega (13) 2k+10}) reduces to%
\begin{equation}
\Omega _{2k+10}^{13}=\frac{1}{\sqrt{3}}\left( -\right) ^{k}\left[ \frac{3}{2}%
\left( k+1\right) a_{2k+2}+\frac{3}{2}ka_{2k}-\frac{1}{2}a_{2k+1}-%
\sum_{m=1}^{k}a_{2k+1-2m}\right]  \label{Omega (13) 2k+10-1}
\end{equation}%
The expression inside the square bracket can be shown to vanish for all
values of $k=0,1,2,3,...$ with the help of the recursion relation\cite%
{Gassem}%
\begin{equation}
\frac{2}{3}a_{n}=\left( n+1\right) a_{n+1}-\left( n-1\right) a_{n-1}
\label{recurrisionRelaaa}
\end{equation}%
To see this we set $n=2k+1$ in the above recursion relation and then
substituting in (\ref{Omega (13) 2k+10-1}) to obtain%
\begin{equation}
\Omega _{2k+10}^{13}=\frac{1}{\sqrt{3}}\left( -\right) ^{k}\left[
3ka_{2k}-\sum_{m=1}^{k}a_{2k+1-2m}\right]  \label{Omega (13) 2k+10-2}
\end{equation}%
Setting $n=2k-1$ in (\ref{recurrisionRelaaa}) and then substituting in (\ref%
{Omega (13) 2k+10-2}), we find%
\begin{equation}
\Omega _{2k+10}^{13}=\frac{1}{\sqrt{3}}\left( -\right) ^{k}\left[ 3\left(
k-1\right) a_{2k-2}-\sum_{m=2}^{k}a_{2k+1-2m}\right]
\label{Omega (13) 2k+10-3}
\end{equation}%
At this point it is not hard to show by explicit substitution that both
expressions inside the square brackets in (\ref{Omega (13) 2k+10-2}) and (%
\ref{Omega (13) 2k+10-3}) vanish for $k=0,1,2$. Now we assume that both
expressions inside the square brackets in (\ref{Omega (13) 2k+10-2}) and (%
\ref{Omega (13) 2k+10-3}) vanish for $k$. We let $k\rightarrow k+1$ in (\ref%
{Omega (13) 2k+10-3}) so that the expression inside the square bracket of (%
\ref{Omega (13) 2k+10-3}) \ becomes%
\begin{equation}
3ka_{2k}-\sum_{m=2}^{k+1}a_{2k+1+2\left( 1-m\right) }
\end{equation}%
Letting $1-m=n$, and then letting $n\rightarrow -n$, the above expression
becomes%
\begin{equation}
3ka_{2k}-\sum_{m=2}^{k+1}a_{2k+1+2\left( 1-m\right)
}=3ka_{2k}-\sum_{n=1}^{k}a_{2k+1-2n}=0
\end{equation}%
where the second equality follows from the fact that the expression inside
the square bracket in (\ref{Omega (13) 2k+10-2}) vanishes for all values of $%
k$. Thus the expression inside the square bracket in (\ref{Omega (13)
2k+10-3}) vanishes identically by mathematical induction and the desired
result follows at once. In fact using this procedure we have checked that
all the identities in (\ref{eqnIdentities Odd0r,s}) are satisfied.

For the case of $m=even$, one can show using similar procedure to the one
used to establish (\ref{eqnIdentities Odd0r,s}) and the following properties
of the $G$ and $\widetilde{N}$ coefficients\cite%
{Gross-Jevicki-II,lousek-Jevicki,ABA,BAA-N,Gassem} 
\begin{eqnarray}
G_{2k\text{ }0}^{13} &=&G_{2k\text{ }0}^{12}\text{, \ \ }G_{2k+1\text{ }%
0}^{13}=-G_{2k+1\text{ }0}^{12} \\
\widetilde{N}_{2k\text{ }0}^{13} &=&\widetilde{N}_{2k\text{ }0}^{12}\text{,
\ \ \ }\widetilde{N}_{2k+1\text{ }0}^{13}=-\widetilde{N}_{2k+1\text{ }0}^{12}
\end{eqnarray}%
that the only none trivial identity in equations (\ref{eqnOmega=0I}) and (%
\ref{eqnOmega=0II}) is%
\begin{equation}
\Omega _{2k\text{ }0}^{11}-\Omega _{2k\text{ }0}^{13}=0
\label{eqn(Omega11-Omega13)2k0}
\end{equation}%
Using equation (\ref{eqnOmegak=0N}), the left hand side of the above
equation becomes%
\begin{eqnarray}
\Omega _{2k\text{ }0}^{11}-\Omega _{2k\text{ }0}^{13} &=&\left( -1\right)
^{k}+k\left( \widetilde{N}_{2k\text{ }0}^{11}-\widetilde{N}_{2k\text{ }%
0}^{13}\right) +\left( 1-6k\right) k\left( G_{2k\text{ }0}^{11}-G_{2k\text{ }%
0}^{13}\right)  \notag \\
&&+\sum_{n=1}^{k-1}\left( 2k-2n\right) \left( -1\right) ^{n}\left( G_{2k-2n%
\text{ }0}^{11}-G_{2k-2n\text{ }0}^{13}\right)
\end{eqnarray}%
Using the values of the $G$ and $\widetilde{N}$ coefficients\cite%
{Gross-Jevicki-II,lousek-Jevicki,ABA,BAA-N,Gassem}%
\begin{eqnarray*}
G_{2k\text{ }0}^{11} &=&\frac{2}{3}\left( -\right) ^{k}\frac{a_{2k}}{2k}%
\text{, \ \ \ \ }G_{2k\text{ }0}^{13}=-\frac{1}{3}\left( -\right) ^{k}\frac{%
a_{2k}}{2k} \\
G_{2k-2n\text{ }0}^{11} &=&\frac{2}{3}\left( -\right) ^{k-n}\frac{a_{2k-2n}}{%
2k-2n}\text{, \ \ \ \ }G_{2k-2n\text{ }0}^{13}=-\frac{1}{3}\left( -\right)
^{k-n}\frac{a_{2k-2n}}{2k-2n} \\
\widetilde{N}_{2k\text{ }0}^{11} &=&-\frac{2}{3}\left( -\right) ^{k}\frac{%
b_{2k}}{2k}\text{, \ \ \ }\widetilde{N}_{2k\text{ }0}^{13}=\frac{1}{3}\left(
-\right) ^{k}\frac{b_{2k}}{2k}
\end{eqnarray*}%
the above expression becomes%
\begin{equation}
\Omega _{2k\text{ }0}^{11}-\Omega _{2k\text{ }0}^{13}=\left( -1\right) ^{k}%
\left[ 1-\frac{1}{2}b_{2k}+\frac{1}{2}\left( 1-6k\right)
a_{2k}+\sum_{n=1}^{k-1}a_{2k-2n}\right]
\end{equation}%
Now to prove (\ref{eqn(Omega11-Omega13)2k0}), we have to show that the
expression inside the square bracket in the above expression vanish for all $%
k$. The proof is very similar to the proof of the identities in (\ref%
{eqnIdentities Odd0r,s}). We can use the mixed recursion relation in (\ref%
{eqnMixed Recab}) to eliminate $b_{2k}$ from the expression inside the
square bracket on the right hand side of the above equation, then with the
help of the recursion relation in (\ref{recurrisionRelaaa}) one can show, by
mathematical induction, that%
\begin{equation}
\left[ 1-\frac{1}{2}b_{2k}+\frac{1}{2}\left( 1-6k\right)
a_{2k}+\sum_{n=1}^{k-1}a_{2k-2n}\right] =0
\end{equation}%
for all values of $k$. Consequently the identity in (\ref%
{eqn(Omega11-Omega13)2k0}) follows at once.

We still need to prove the identity in (\ref{eqn GeneralIDforOmegars,mk}).
Unfortunately this identity is quite difficult to prove; the difficulty
arises because of the finite sums over the $G$ and $\widetilde{N}$
coefficients. There is no obvious way of proving this identity in a clean
way other than by brute force. However let us start by working out few
specific cases. We first consider $\Omega _{11}^{rs}$. letting $m=k=1$ in (%
\ref{eqnOmegars;km}) we have%
\begin{equation}
\Omega _{11}^{rs}=-\delta ^{rs}-G_{11}^{rs}+2\widetilde{N}_{11}^{rs}
\label{eqnOmegars-11}
\end{equation}%
For $r=s$; we have $G_{11}^{rr}=-5/3^{3}$ and $\widetilde{N}%
_{11}^{rs}=11/3^{3}$, so the right hand side of the above equation vanish
for $r=s$. Thus $\Omega _{11}^{rr}=0$ for $r=1,2,3$. For $r=1$, $s=2,3$, we
have $G_{11}^{12}=G_{11}^{13}=2^{4}/3^{3}$, $\widetilde{N}_{11}^{12}=%
\widetilde{N}_{11}^{13}=2^{3}/3^{3}$. Using these values we see that $%
-\delta ^{rs}-G_{11}^{rs}+2\widetilde{N}_{11}^{rs}=0$ for $r=1$, $s=2,3$ and
so we have $\Omega _{11}^{12}=\Omega _{11}^{13}=0$. The cyclic symmetries of 
$G_{nm}^{rs}$ and $\widetilde{N}_{nm}^{rs}$ now imply that the right hand
side of (\ref{eqnOmegars-11}) also vanish for $\left( r,s\right) =\left(
2,3\right) $, $\left( 3,1\right) $, $\left( 3,2\right) $ and $\left(
2,1\right) $. Thus we have established that $\Omega _{11}^{rs}=0$ for all $r$
and $s$. We will work out few more case; however in what follows we shall
not refer to the exact equations explicitly for the $G_{nm}^{rs}$ and $%
\widetilde{N}_{nm}^{rs}$ coefficients\cite%
{Gross-Jevicki-II,lousek-Jevicki,ABA,BAA-N,Gassem} and it will not be much
of a task for the reader to look them up. Let us next consider the case when 
$k+m=odd$. For $k=odd$ and $\ m=even$ equation (\ref{eqnOmegars;km}) yields%
\begin{eqnarray}
\Omega _{2m\text{ }2k+1}^{rs} &=&\sum_{n=1}^{m-1}2nG_{2n\text{ }%
2k+1}^{rs}\left( -1\right) ^{m-n}+\left( 1-6m\right) mG_{2m\text{ }%
2k+1}^{rs}+\sum_{n=1}^{k}2m\times   \notag \\
&&\widetilde{N}_{2m\text{ }2n-1}^{rs}\left( -1\right) ^{k-n+1}+m\widetilde{N}%
_{2m\text{ }2k+1}^{rs}\left( 1+3\left( 2k+1\right) \right) 
\label{eqnOmegars2m-2k+1}
\end{eqnarray}%
First consider $r=s$; in this case $G_{n+m=ood}^{rr}=\widetilde{N}%
_{n+m=odd}^{rr}=0$ and the right hand side vanish; hence $\Omega _{2m\text{ }%
2k+1}^{rr}=0$. Likewise one can show that $\Omega _{2m+1\text{ }2k}^{rr}=0$.
Next we consider $\Omega _{2m\text{ }2k+1}^{12}=\Omega _{2m\text{ }%
2k+1}^{23}=\Omega _{2m\text{ }2k+1}^{31}$, where the equality here follows
from the cyclic symmetry of the $G_{nm}^{rs}$ and $\widetilde{N}_{nm}^{rs}$
coefficients. Thus it suffice in this case to show that $\Omega _{2m\text{ }%
2k+1}^{12}=0$. To prove this put $r=1$, $s=2$ in equation (\ref%
{eqnOmegars2m-2k+1})%
\begin{eqnarray}
\Omega _{2m\text{ }2k+1}^{12} &=&\left( -1\right) ^{m}\sum_{n=1}^{m-1}2nG_{2n%
\text{ }2k+1}^{12}\left( -1\right) ^{n}+\left( 1-6m\right) mG_{2m\text{ }%
2k+1}^{12}-\left( -1\right) ^{k}  \notag \\
&&\sum_{n=0}^{k-1}2m\widetilde{N}_{2m\text{ }2n+1}^{12}\left( -1\right)
^{n+1}+m\widetilde{N}_{2m\text{ }2k+1}^{12}\left( 1+3\left( 2k+1\right)
\right)   \label{eqnOmega12-2m-2k+1}
\end{eqnarray}%
At this point we can use the explicit values \cite%
{Gross-Jevicki-II,Hlousek-Jevicki,ABA,BAA-N,Gassem} of the $G_{2n\text{ }%
2k+1}^{12}$, $G_{2m\text{ }2k+1}^{12}$, $\widetilde{N}_{2m\text{ }2n-1}^{12}$
and $\widetilde{N}_{2m\text{ }2k+1}^{12}$coefficients and carry the
calculation to the end however this will take too much space. So at this
stage let us work out specific cases. For $m=1$, $k=0$, the above expression
becomes%
\begin{equation*}
\Omega _{2\text{ }1}^{12}=-5G_{2\text{ }1}^{12}+4\widetilde{N}_{2\text{ }%
1}^{12}
\end{equation*}%
Since $G_{2\text{ }1}^{12}=2^{5}/3^{4}\sqrt{3}$, $\widetilde{N}_{2\text{ }%
1}^{12}=-2^{3}\cdot 5/3^{4}\sqrt{3}$, the right hand side of the above
equation is identically zero. Hence, we have established that $\Omega _{2%
\text{ }1}^{12}=0$. For $m=1$, $k=1$, equation (\ref{eqnOmega12-2m-2k+1})
gives%
\begin{equation}
\Omega _{2\text{ }3}^{12}=-5G_{2\text{ }3}^{12}-2\widetilde{N}_{2\text{ }%
1}^{12}+10\widetilde{N}_{2\text{ }3}^{12}
\end{equation}%
The $G$ and $\widetilde{N}$ are given by\cite%
{Gross-Jevicki-II,Hlousek-Jevicki,ABA,BAA-N,Gassem} $G_{2\text{ }%
3}^{12}=-4\cdot 2^{3}\cdot 5/3^{6}\sqrt{3}$, $\widetilde{N}_{2\text{ }%
1}^{12}=-2^{3}\cdot 5/3^{4}\sqrt{3}$, $\widetilde{N}_{2\text{ }%
3}^{12}=-19\cdot 2^{3}/3^{6}\sqrt{3}$. Substituting these values in the
above expression yields $\Omega _{2\text{ }3}^{12}=0$.

From these examples, we see that the difficulty involved in constructing a
general proof of these identities. However, we still can describe the steps
involved in the proof and give some of the details. Using the $G_{nm}^{rs}$
and $\widetilde{N}_{nm}^{rs}$ coefficients, one can show that the only
nontrivial identities in (\ref{eqn GeneralIDforOmegars,mk}) are the
following four%
\begin{equation}
\Omega _{2m\text{ }2k+1}^{12}=\Omega _{2m+1\text{ }2k}^{12}=\Omega _{2m\text{
}2k}^{11}=\Omega _{2m+1\text{ }2k+1}^{11}
\end{equation}%
and all the other identities are either trivially satisfied or can be
deduced from these four identities. To illustrate the proof of the above
identities we consider $\Omega _{2m\text{ }2k+1}^{12}$. Substituting the
explicit values\cite{Gross-Jevicki-II,ABA,BAA-N,Gassem,Hlousek-Jevicki} for
the $G$ and $\widetilde{N}$ coefficients%
\begin{eqnarray}
&&G_{2n2m+1}^{12}  \notag \\
&=&\frac{\left( -\right) ^{n+m}}{2\sqrt{3}}\left[ \frac{%
a_{2n}b_{2m+1}-b_{2n}a_{2m+1}}{\left( 2n\right) +\left( 2m+1\right) }+\frac{%
a_{2n}b_{2m+1}+b_{2n}a_{2m+1}}{\left( 2n\right) -\left( 2m+1\right) }\right] 
\\
&&\widetilde{N}_{2n2m+1}^{12}  \notag \\
&=&\frac{\left( -1\right) ^{n+m}}{2\sqrt{3}}\left[ \frac{%
b_{2n}a_{2m+1}-a_{2n}b_{2m+1}}{2n+\left( 2m+1\right) }+\frac{%
b_{2n}a_{2m+1}+a_{2n}b_{2m+1}}{2n-\left( 2m+1\right) }\right] 
\end{eqnarray}%
into equation (\ref{eqnOmega12-2m-2k+1}) , and skipping some algebra we
obtain

\begin{eqnarray*}
&&\frac{2\sqrt{3}}{\left( -\right) ^{m+k}}\Omega _{2m\text{ }2k+1}^{12} \\
&=&\sum_{n=0}^{m-1}2n\left[ \frac{a_{2n}b_{2k+1}-b_{2n}a_{2k+1}}{\left(
2n\right) +\left( 2k+1\right) }+\frac{a_{2n}b_{2k+1}+b_{2n}a_{2k+1}}{\left(
2n\right) -\left( 2k+1\right) }\right] \\
&&+\sum_{n=0}^{k-1}2m\left[ \frac{b_{2m}a_{2n+1}-a_{2m}b_{2n+1}}{2m+\left(
2n+1\right) }+\frac{b_{2m}a_{2n+1}+a_{2m}b_{2n+1}}{2m-\left( 2n+1\right) }%
\right] \\
&&-\frac{2m}{2k-2m+1}\left(
4a_{2m}b_{2k+1}+b_{2m}a_{2k+1}+6ka_{2m}b_{2k+1}-6ma_{2m}b_{2k+1}\right)
\allowbreak
\end{eqnarray*}%
At this stage we have checked by explicit substitution of the $a$'s and the $%
b$'s that the right hand side vanish for the first few low values of $m$ and 
$k$. Theses consistency checks are important to ensure that we are on the
right track. Now we use the mixed recursion relation in (\ref{eqnMixed Recab}%
) to eliminate the $b$' in favor of the $a$'s; hence using (\ref{eqnMixed
Recab}), the expression becomes

\begin{eqnarray}
&&\frac{2\sqrt{3}}{\left( -\right) ^{m+k}}\Omega _{2m\text{ }%
2k+1}^{12}=\sum_{n=0}^{m-1}\frac{-1}{4k^{2}+4k-4n^{2}+1}\{6na_{2k+1}a_{2n+1}
\notag \\
&&-6na_{2k+1}a_{2n-1}-24n^{2}a_{2n}a_{2k+2}+12n^{2}a_{2k+1}a_{2n-1}+12n^{2}a_{2k+1}a_{2n+1}
\notag \\
&&-24kn^{2}a_{2k}a_{2n}-12kna_{2k+1}a_{2n-1}+12kna_{2k+1}a_{2n+1}  \notag \\
&&-24kn^{2}a_{2n}a_{2k+2}+24kn^{2}a_{2k+1}a_{2n-1}+24kn^{2}a_{2k+1}a_{2n+1}\}
\notag \\
+ &&\sum_{n=0}^{k-1}\frac{1}{-4m^{2}+4n^{2}+4n+1}%
\{12ma_{2m}a_{2n+2}-12ma_{2m}a_{2n+1}  \notag \\
&&+48m^{3}a_{2m}a_{2n+1}+12m^{2}a_{2m-1}a_{2n+1}-12m^{2}a_{2m+1}a_{2n+1} 
\notag \\
&&-24m^{3}a_{2m-1}a_{2n+1}-24m^{3}a_{2m+1}a_{2n+1}  \notag \\
&&-48mna_{2m}a_{2n+1}+36mna_{2m}a_{2n+2}+24mn^{2}a_{2m}a_{2n}  \notag \\
&&-48mn^{2}a_{2m}a_{2n+1}+24mn^{2}a_{2m}a_{2n+2}+12mna_{2m}a_{2n}\}  \notag
\\
&&+2\frac{m}{2k-2m+1}\{12a_{2m}a_{2k+2}-12a_{2m}a_{2k+1}-3a_{2k+1}a_{2n+1} 
\notag \\
&&+3a_{2k+1}a_{2n+2}-42ka_{2m}a_{2k+1}+30ka_{2m}a_{2k+2}+18ma_{2m}a_{2k+1} 
\notag \\
&&-18ma_{2m}a_{2k+2}+3na_{2n}a_{2k+1}+18k^{2}a_{2k}a_{2m}-6na_{2k+1}a_{2n+1}
\notag \\
&&+3na_{2k+1}a_{2n+2}-36k^{2}a_{2m}a_{2k+1}  \notag \\
&&+18k^{2}a_{2m}a_{2k+2}+12ka_{2k}a_{2m}+36kma_{2m}a_{2k+1}-18kma_{2m}a_{2k+2}
\notag \\
&&-18kma_{2k}a_{2m}\}  \label{eqnOmega122mak+1Factor}
\end{eqnarray}%
At this point we have checked by explicit substitution that the right hand
side is zero for the first few values of $m$ and $k$. So we will prove this
identity by mathematical induction. Let us assume the right hand side of the
above equation vanish for given $m$ and $k$, then for $m\rightarrow m+1$ and 
$k\rightarrow k+1$, the right hand side (RHS) reads

\begin{eqnarray*}
&&\text{RHS} \\
&=&\sum_{n=0}^{m}\frac{-6n}{4k^{2}+12k-4n^{2}+9}\{\allowbreak
3a_{2k+3}a_{2n+1}-3a_{2k+3}a_{2n-1}-8na_{2n}a_{2k+4} \\
&&-\allowbreak 2ka_{2k+3}a_{2n-1}+2ka_{2k+3}a_{2n+1}+6na_{2k+3}\allowbreak
a_{2n-1}+6na_{2k+3}a_{2n+1} \\
&&-4kna_{2n}a_{2k+2}-\allowbreak
4kna_{2n}a_{2k+4}+4kna_{2k+3}a_{2n-1}+4kna_{2k+3}a_{2n+1}\} \\
&&+\sum_{n=0}^{k}\frac{1}{-4\left( m+1\right) ^{2}+4n^{2}+4n+1}\{ \\
&&12a_{2m+2}\left( m+1\right) \left(
4a_{2n+1}m^{2}+8a_{2n+1}m+3a_{2n+1}+a_{2n+2}\right) \\
&&-12a_{2n+1}\left( m+1\right) ^{2}\left(
2ma_{2m+1}+2ma_{2m+3}+a_{2m+1}+3a_{2m+3}\right) \\
&&-12na_{2m+2}\left( m+1\right) \left(
4na_{2n+1}-2na_{2n}+4a_{2n+1}-3a_{2n+2}\right) \\
&&+12n\left( 2na_{2n+2}+a_{2n}\right) a_{2m+2}\left( m+1\right) \} \\
&&+\frac{2\left( m+1\right) }{2\left( k+1\right) -2\left( m+1\right) +1}%
\{12a_{2k+4}a_{2m+2}-3a_{2k+3}a_{2n+1} \\
&&-12a_{2k+3}a_{2m+2}+3a_{2k+3}a_{2n+2}+3na_{2n}a_{2k+3}-18a_{2k+4}a_{2m+2}-
\\
&&24a_{2k+3}a_{2m+2}-42ka_{2k+3}a_{2m+2}+30ka_{2k+4}a_{2m+2}+18ma_{2k+3}a_{2m+2}
\\
&&-18ma_{2k+4}a_{2m+2}+18a_{2k+2}a_{2m+2}-36a_{2k+3}a_{2m+2}+36ka_{2k+2}a_{2m+2}
\\
&&-72ka_{2k+3}a_{2m+2}-6na_{2k+3}a_{2n+1}+3na_{2k+3}a_{2n+2}+18k^{2}a_{2k+2}a_{2m+2}
\\
&&-36k^{2}a_{2k+3}a_{2m+2}+6a_{2m+2}\left( k+1\right) \times \\
&&\left(
3ka_{2k+4}-3ma_{2k+2}+6ma_{2k+3}-3ma_{2k+4}-a_{2k+2}+6a_{2k+3}\right) \}
\end{eqnarray*}%
This expression reduces to the right hand side of (\ref%
{eqnOmega122mak+1Factor}). To see this one needs only to use the recursion
relations in (\ref{recurrisionRelaaa}), to reduces the indices to those
appearing in (\ref{eqnOmega122mak+1Factor}). We have checked that this is
indeed true, however the calculation is quite messy to include here,
otherwise straight forward. Thus the right hand side of (\ref%
{eqnOmega122mak+1Factor}) vanish identically by induction for all values of $%
m$ and $k$. Likewise one can prove the remaining identities; the procedure
is straight forward but it takes countless number of pages and so we shall
not include any more details here.

\section{\protect\bigskip BRST invariance}

The full Ward-like identity now reads%
\begin{equation}
\left[ L_{m}^{x+\phi ,r}+\sum_{s=1}^{3}\sum_{n=0}^{\infty }m\widetilde{N}%
_{mn}^{rs}L_{-n}^{x+\phi ,s}\right] |V_{HS}^{x+\phi }>=0\text{, \ \ }%
m=1,2,...  \label{eqnWard-like-Ide}
\end{equation}%
We have seen that the coordinate and the ghost anomaly cancel; thus the
above equation contain no anomaly term. The $K_{m}$ invariance of the comma
interaction three vertex follows at once from the Ward-like identity.
Summing over the string index $r$, we have%
\begin{equation}
\sum_{r=1}^{3}\left[ L_{m}^{x+\phi ,r}+\sum_{s=1}^{3}\sum_{n=0}^{\infty }m%
\widetilde{N}_{mn}^{rs}L_{-n}^{x+\phi ,s}\right] |V_{HS}^{x+\phi }>=0\text{,
\ \ }m=1,2,...
\end{equation}%
Using the identity $\sum_{r=1}^{3}m\widetilde{N}_{mn}^{rs}=\left( -1\right)
^{m+1}\delta _{mn}$, which can be established by contour integration, and
then renaming the dummy index $s$ as $r$, the above equation reduces to%
\begin{equation}
\sum_{r=1}^{3}\left[ L_{m}^{x+\phi ,r}-\left( -1\right) ^{m}L_{-m}^{x+\phi
,r}\right] |V_{HS}^{x+\phi }>=0\text{, \ \ }m=1,2,...
\end{equation}%
The expression inside the square bracket in \ the above equation is by
definition the Virasoro generator $K_{m}$. Thus $|V_{HS}^{x+\phi }>$ is
invariant under the subgroup of conformal transformations, generated by the
Virasoro generators%
\begin{equation}
K_{m}=\sum_{r=1}^{3}\left[ L_{m}^{x+\phi ,r}-\left( -1\right)
^{m}L_{-m}^{x+\phi ,r}\right]
\end{equation}

\bigskip To complete the proof of equivalence, we still need to show the $%
BRST$ \ ($Q$) invariance of the comma three vertex. Unfortunately a direct
proof that follows from equation (\ref{eqnWard-like-Ide}) is quite
cumbersome due to the presence of the $\frac{1}{2}L_{m}^{\phi ,r}$ term in
the definition of the $BRST$ charge. Thus we need to evaluate the action of
the $BRST$ on the comma three vertex directly. Recall that the total three
string $BRST$ charge is the sum of the $BRST$ charges corresponding to the
individual strings; that is,%
\begin{equation}
Q=\sum_{r=1}^{3}Q^{r}
\end{equation}%
where%
\begin{eqnarray}
Q^{r} &=&\sum_{m=1}^{\infty }\left[ c_{-m}^{r}\left( L_{m}^{x,r}+\frac{1}{2}%
L_{m}^{\phi ,r}\right) +\left( L_{-m}^{x,r}+\frac{1}{2}L_{-m}^{\phi
,r}\right) c_{m}^{r}\right]  \notag \\
&&+\left( L_{0}^{x,r}+\frac{1}{2}L_{0}^{\phi ,r}-1\right) c_{0}^{r}
\end{eqnarray}%
To evaluate the action of $BRST$ charge on the full comma three vertex we
use the $c$-overlaps satisfied by the half string ghost vertex\cite%
{Gross-Jevicki-II,Abdu-BordesN-II} 
\begin{equation}
\left[ c_{m}^{r}-\sum_{s=1}^{3}\sum_{n=1}^{\infty }\widetilde{N}%
_{mn}^{rs}c_{-n}^{s}\right] |V_{HS}^{\phi }>=0  \label{eqnCoverlonHSV}
\end{equation}%
With the help of (\ref{eqnCoverlonHSV}), the proof of the $BRST$ invariance
follows along the same lines of references\cite%
{Gross-Jevicki-II,Abdu-BordesN-II}. Thus when acting with the $BRST$ charge
on the full half string vertex, the operator parts cancel, leaving%
\begin{eqnarray}
Q|V_{HS}^{x+\phi } &>&=\sum_{r=1}^{3}\sum_{m=1}^{\infty }c_{-m}^{r}\left[
\left( \kappa _{m}^{x,r}+\frac{1}{2}\kappa _{m}^{\phi ,r}\right) \right. -%
\frac{1}{2}\times  \notag \\
&&\left. \left( m^{2}\widetilde{N}_{0m}^{rr}+\sum_{k=1}^{m-1}\left(
m+k\right) \left( m-k\right) \widetilde{N}_{m\text{ }m-k}^{rr}\right) \right]
|V_{HS}^{x+\phi }>
\end{eqnarray}%
The expression inside the second parenthesis can be computed easily and it
is found to be the anomaly of the fermionic ghost\cite%
{Gross-Jevicki-II,Abdu-BordesN-II}. Thus the coefficient of $c_{-m}^{r}$ is%
\begin{equation}
\kappa _{m}^{x,r}+\frac{1}{2}\kappa _{m}^{\phi ,r}+\frac{1}{2}\kappa
_{m}^{c,r}=0
\end{equation}%
where in obtaining the above result we used the fact that the coordinate
anomaly is the negative of the ghost anomaly regardless of the ghost
representation. This result is the final step in the proof of equivalence.

\appendix{}

\section{Sums of the first type}

Generalizations of the above sums%
\begin{equation}
O_{n=2k}^{u(q,p)}=\sum_{m=2l+1=1}^{\infty }\frac{u_{m}^{q/p}}{n+m}\text{ , }%
n\geq 0  \label{eqnsumoverodd1}
\end{equation}%
\begin{equation}
E_{n=2k+1}^{u(q,p)}=\sum_{m=2l=0}^{\infty }\frac{u_{m}^{q/p}}{n+m}\text{ , }%
n>0  \label{eqnsumovereven1}
\end{equation}%
where $u_{m}^{q/p}$ are the Taylor modes appearing in the expansion%
\begin{equation}
\left( \frac{1+z}{1-z}\right) ^{q/p}=\sum_{m=0}^{\infty }u_{n}^{q/p}z^{n}
\label{equationdefofu(q/p)}
\end{equation}%
Mathematical induction leads to%
\begin{equation}
E_{-n=-(2k+1)}^{u(q,p)}=-\cos \left( \frac{\pi q}{p}\right)
E_{n=2k+1}^{u(q,p)}\text{, \ \ \ }n>0  \label{eqnSumOnegative n=odd}
\end{equation}%
and%
\begin{equation}
O_{-n=-2k}^{u(q,p)}=-\cos \left( \frac{\pi q}{p}\right) O_{n=2k}^{u(q,p)}%
\text{, \ \ \ }n>0  \label{eqnSumOnegative n=even}
\end{equation}%
respectively. So far we have evaluated the sums defined in (\ref%
{eqnsumoverodd1}) and (\ref{eqnsumovereven1}) under the restriction $n+m=odd$%
; now we would like to relax this restriction, which brings us to the sums
of the second type.

\section{Sums of the second type}

Sums of the second type are defined by%
\begin{equation}
S_{n=2k+1}^{u(q,p)}\equiv O_{n=2k+1}^{u(q,p)}=\sum_{m=2l+1=1}^{\infty }\frac{%
u_{m}^{q/p}}{n+m}  \label{eqnSUMeveneven}
\end{equation}%
\begin{equation}
S_{n=2k}^{u(q,p)}\equiv E_{n=2k}^{u(q,p)}=\sum_{m=2l=0}^{\infty }\frac{%
u_{m}^{q/p}}{n+m}  \label{eqnSUModdodd}
\end{equation}%
\begin{eqnarray}
S_{n}^{u(q/p)} &=&\frac{p}{2q}\left[ \frac{q}{p}\left( \beta \left( 1-\frac{q%
}{p}\right) +\beta \left( 1+\frac{q}{p}\right) \right) \right] u_{n}^{q/p} 
\notag \\
&&+\frac{p}{2q}\sum_{m=0}^{n-1}\left( -\right) ^{m}\frac{%
u_{m}^{q/p}u_{n-m-1}^{q/p}}{m+1}  \label{eqn.Gen.SnU}
\end{eqnarray}

\section{Sums of the third type}

\begin{equation}
\overset{\sim }{O}_{n=2k}^{u(q,p)}=\sum_{m=2l+1=1}^{\infty }\frac{u_{m}^{q/p}%
}{\left( n+m\right) ^{2}}  \label{eqnOteldan=even}
\end{equation}%
\begin{equation}
\overset{\sim }{E}_{n=2k+1}^{u(q,p)}=\sum_{m=2l=0}^{\infty }\frac{u_{m}^{q/p}%
}{\left( n+m\right) ^{2}}  \label{eqnEteldan=odd}
\end{equation}

\begin{eqnarray}
\overset{\sim }{E}_{n=odd=1}^{u(p,q)} &=&-\frac{1}{2}\left( \frac{q}{p}%
\right) \frac{\pi }{\sin \left( \pi q/p\right) }\left\{ \left[ 2\psi \left( 
\frac{q}{p}\right) +\frac{1}{\left( q/p\right) }-2\psi \left( 1\right)
-2+2\ln 2\right] \right.  \notag \\
&&+\left. \cos \left( \pi q/p\right) \left[ \psi \left( \frac{q}{2p}+\frac{1%
}{2}\right) -\psi \left( \frac{q}{2p}\right) -\frac{1}{\left( q/p\right) }%
\right] \right\}  \label{EqnE(n=odd)}
\end{eqnarray}

\begin{eqnarray}
\overset{\sim }{S}_{n}^{u(q,p)} &=&\overset{\sim }{E}_{n}^{u(q,p)}\text{, \
\ \ \ }n=2k+1>0  \notag \\
\overset{\sim }{S}_{n}^{u(q,p)} &=&\overset{\sim }{O}_{n}^{u(q,p)}\text{, \
\ }n=2k\geqslant 0\text{ }  \label{eqnSTelda}
\end{eqnarray}%
and%
\begin{eqnarray}
\overline{S}_{n}^{u(q,p)} &=&E_{n}^{u(q,p)}\text{, \ \ \ \ }n=2k+1>0  \notag
\\
\overline{S}_{n}^{u(q,p)} &=&O_{n}^{u(q,p)}\text{, \ \ }n=2k\geqslant 0
\label{eqnSbar}
\end{eqnarray}%
\begin{eqnarray}
{}\widetilde{S}_{n}^{u(q/p)} &=&\left[ \widetilde{S}_{1}^{u(q/p)}-\frac{%
2q\pi }{2p\sin \left( \pi q/p\right) }\right] \frac{pu_{n}^{q/p}}{2q}-\frac{%
\pi }{2\sin \left( \pi q/p\right) }\sum_{k=1}^{n}\frac{\left( -\right) ^{k}}{%
k}u_{k}^{q/p}u_{n-k}^{q/p}  \notag \\
&&-\frac{\pi }{2}\tan \left( \pi \frac{q}{2p}\right) \frac{p}{2q}%
\sum_{k=0}^{n-1}\frac{\left( -\right) ^{k}}{k+1}u_{k}^{q/p}u_{n-1-k}^{q/p}%
\text{ \ \ , }n>0  \label{eqnEXPValOfSteln}
\end{eqnarray}%
This result holds for all integer values of $n\geq 1$.

\begin{equation}
\widetilde{S}_{-n}^{u(q,p)}=\cos \left( \pi \frac{q}{p}\right) \widetilde{S}%
_{n}^{u(q,p)}+\left[ 1+\cos \left( \pi \frac{q}{p}\right) \right] \overline{S%
}_{0}^{u(q,p)}S_{n}^{u(q,p)}  \label{eqnSUM-S-nTel}
\end{equation}%
\begin{equation*}
\overline{S}_{0}^{u(q,p)}=\frac{1}{2}\pi \tan \left( \pi \frac{q}{2p}\right)
\end{equation*}

\bigskip\ \

\bigskip

\bigskip

\end{document}